\newcommand{\CLASSINPUTtoptextmargin}{2.36cm}
\documentclass[12pt,draftclsnofoot,onecolumn]{IEEEtran}

\usepackage{amsmath, mathrsfs, amsthm}
\usepackage{amsfonts}
\usepackage{amssymb}
\usepackage{graphicx}
\usepackage{color}
\usepackage{multirow}
\usepackage{graphicx}
\usepackage{stmaryrd}
\usepackage{flafter}
\usepackage{tabularx}

\usepackage{algorithm}
\usepackage{algpseudocode}
\usepackage{pifont}
\usepackage{varwidth}

\long\def\comment#1{}

\newcommand{\RR}{\mbox{\bb R}}

\newcommand{\Phim}{\hbox{\boldmath$\Phi$}}

\newcommand{\transp}{{\sf T}}

\usepackage{mathbbol}

\def\mindex#1{\index{#1}}

\def\sq{\hbox{\rlap{$\sqcap$}$\sqcup$}}
\def\qed{\ifmmode\sq\else{\unskip\nobreak\hfil
\penalty50\hskip1em\null\nobreak\hfil\sq
\parfillskip=0pt\finalhyphendemerits=0\endgraf}\fi\medskip}

\long\def\defbox#1{\framebox[.9\hsize][c]{\parbox{.85\hsize}{%
\parindent=0pt
\baselineskip=12pt plus .1pt      % STYLE
\parskip=6pt plus 1.5pt minus 1pt % CHANGES
 #1}}}

\long\def\beginbox#1\endbox{\subsection*{}%
\hbox{\hspace{.05\hsize}\defbox{\medskip#1\bigskip}}%
\subsection*{}}

\def\endbox{}

\newsavebox{\junk}
\savebox{\junk}[1.6mm]{\hbox{$|\!|\!|$}}

\def\limsup{\mathop{\rm lim\ sup}}

\def\argmin{\mathop{\rm arg\, min}}

\def\argmax{\mathop{\rm arg\, max}}

\def\bC{{\mathbb C}}

\def\bE{{\mathbb E}}

\def\bR{{\mathbb R}}

\def\bfI{{\bf I}}

\def\bfb{{\bf b}}
\def\bfc{{\bf c}}

\def\bfh{{\bf h}}

\def\bfq{{\bf q}}

\def\bfs{{\bf s}}
\def\bft{{\bf t}}

\def\bfv{{\bf v}}
\def\bfw{{\bf w}}

\def\bfz{{\bf z}}

\def\scrP{{\mathscr{P}}}

\def\scrR{{\mathscr{R}}}

\def\scrT{{\mathscr{T}}}

\def\bfmath#1{{\mathchoice{\mbox{\boldmath$#1$}}%
{\mbox{\boldmath$#1$}}%
{\mbox{\boldmath$\scriptstyle#1$}}%
{\mbox{\boldmath$\scriptscriptstyle#1$}}}}

\def\bfmY{\bfmath{Y}}

\def\bfmhhaY{\bfmath{\hhaY}} %\widehat{\widehat{Y}}}}
\def\bfmhhaY{\hbox to 0pt{$\widehat{\bfmY}$\hss}\widehat{\phantom{\raise 1.25pt\hbox{$\bfmY$}}}}

\def\til={{\widetilde =}}

\def\clA{{\cal A}}
\def\clB{{\cal B}}
\def\clC{{\cal C}}

\def\clN{{\cal N}}

\def\clS{{\cal S}}

\def\clY{{\cal Y}}
\def\clZ{{\cal Z}}

 \def\FRAC#1#2#3{\genfrac{}{}{}{#1}{#2}{#3}}

\def\ddtp{{\mathchoice{\FRAC{1}{d^{\hbox to 2pt{\rm\tiny +\hss}}}{dt}}%
{\FRAC{1}{d^{\hbox to 2pt{\rm\tiny +\hss}}}{dt}}%
{\FRAC{3}{d^{\hbox to 2pt{\rm\tiny +\hss}}}{dt}}%
{\FRAC{3}{d^{\hbox to 2pt{\rm\tiny +\hss}}}{dt}}}}

\def\average#1,#2,{{1\over #2} \sum_{#1}^{#2}}

\def\eye(#1){{\bf(#1)}\quad}

\newtheorem{definition}{{\bf Definition}}

\newtheorem{remark}{{\bf Remark}}

\def\eq#1/{(\ref{e:#1})}

\newcommand{\beqn}[1]{\notes{#1}%
\begin{eqnarray} \elabel{#1}}

\newcommand{\eeqn}{\end{eqnarray} }

\newcommand{\beq}[1]{\notes{#1}%
\begin{equation}\elabel{#1}}

\newcommand{\eeq}{\end{equation}}

\def\bdes{\begin{description}}
\def\edes{\end{description}}

\newcounter{rmnum}

\newcounter{anum}

{\end{list}}

\def\ass(#1:#2){(#1\ref{#1:#2})}

\def\ritem#1{
\item[{\sf \ass(\current_model:#1)}]
}

\newenvironment{recall-ass}[1]{%
\begin{description}
\def\current_model{#1}}{
\end{description}
}

\usepackage[numbers,sort&compress]{natbib}

\usepackage{tikz}
\usepackage{pgfplots,tikz-3dplot}
\usepackage{pgfplotstable}%used in bar charts
\pgfplotsset{compat=newest}
\usetikzlibrary{patterns}
\usetikzlibrary{positioning}
\usetikzlibrary{datavisualization}
\usetikzlibrary{datavisualization.formats.functions}
\usetikzlibrary{backgrounds}
\usetikzlibrary{shapes,snakes}
\def\cg{{\clC\clN}} 
\begin{document}
%\title{low complexity detection in  MIMO using AMP with Gaussian prior}
\title{DRL-Based QoS-Aware Resource Allocation Scheme for Coexistence of Licensed and Unlicensed Users in LTE and Beyond}

\author{\IEEEauthorblockN{Mahdi Nouri Boroujerdi\IEEEauthorrefmark{1}, Mohammad Akbari\IEEEauthorrefmark{2}, Roghayeh Joda\IEEEauthorrefmark{2}, Mohammad Ali Maddah-Ali\IEEEauthorrefmark{1}, Babak Hossein Khalaj\IEEEauthorrefmark{1}}\\
%	
%\IEEEauthorblockA{\IEEEauthorrefmark{1}School of Electrical and Computer Engineering,
%University of Tehran,
%Tehran, Iran \Red{Sharif?}\\
%Email: mhd.nouri@ut.ac.ir}\\
\IEEEauthorblockA{\IEEEauthorrefmark{1}Department of Electrical Engineering, Sharif University of Technology, Tehran, Iran\\
	Email: \{ma.nouri, maddah\_ali, khalaj\}@sharif.edu}\\
\IEEEauthorblockA{\IEEEauthorrefmark{2}Department of Communication Technologies, ICT Research Center (ITRC), Tehran, Iran\\
Email:\{m.akbari, r.joda\}@itrc.ac.ir}\\

}




\maketitle

\begin{abstract}
In this paper, we employ deep reinforcement learning to develop a novel radio resource allocation and packet scheduling scheme for  different Quality of Service (QoS) requirements applicable to LTE-advanced and 5G networks. In addition, regarding the scarcity of spectrum in below 6\,GHz bands, the proposed algorithm dynamically allocates the resource blocks (RBs) to licensed users in a way to mostly preserve the continuity of unallocated RBs. This would improve the efficiency of communication among the unlicensed entities by increasing the chance of uninterrupted communication and reducing the load of coordination overheads. The optimization problem is formulated as a Markov Decision Process (MDP), observing the entire queue of the demands, where failing to meet QoS constraints penalizes the goal with a multiplicative factor. Furthermore, a notion of continuity for unallocated resources is taken into account as an additive term in the objective function. Considering the variations in both channel coefficients and users' requests, we utilize a deep reinforcement learning algorithm as an online and numerically efficient approach to solve the MDP. Numerical results show that the proposed method achieves higher average spectral efficiency, while considering delay budget and packet loss ratio, compared to the conventional greedy min-delay and max-throughput schemes, in which a fixed part of the spectrum is forced to be vacant for unlicensed entities.

\end{abstract}


% \fi
\IEEEpeerreviewmaketitle

\section{Introduction}
\label{sec:intro}

%SON, in general, has many potential applications of such scenario \cite{SON_smallcell13, SON_bandit17, SON_MAB16}.
 With rapidly growing demand for mobile wide-band radio access, developing bandwidth-efficient strategies has the highest importance.
 Allowing different operators to share the same bandwidth, Dynamic Spectrum Sharing (DSS) is known as one of the most promising 5G enablers, alleviating the lack of spectral efficiency \cite{DSS_cognitive_survey}.
% 	 Typically, users compete to find spectrum holes in unlicensed bands (ref). For licensed bands, this is true as long as the licensed users are not compromised.
% Resource allocation strategies in LTE are classified into different categories in the literature. Some of the more famous categories are: Channel unaware, channel aware/QoS unaware, channel and QoS aware . The first group includes simple schemes such as equal throughput or round robin scheduling. The second type uses Channel Quality Index (CQI) to achieve higher system's throughput. Maximum throughput is among these schemes. The last category takes into account some QoS requirements, such as bit rate or delay constraint, by the users as well. Many of the works in the literature fall in this category \cite{pkt_schedule_LTE13}.
 LTE-Unlicensed (LTE-U) is a good example of spectrum sharing, where the main idea is to offload a fraction of LTE traffic to unlicensed bands in order to increase the overall throughput of a typical LTE system, while minimizing the degradation in performance of WiFi users \cite{LTE-UvsLAA_19, LTE-U_sensing20}.\\
   On the other hand, efficient resource allocation and packet scheduling scheme is a keystone in designing a bandwidth-efficient strategy. Classic resource allocation and packet scheduling schemes in LTE are based on simple heuristics such as round robin scheduling or equal throughput for all users in the network. More advanced schemes include some sort of feedback from the user side as well, such as channel coefficients, to optimize a predefined metric, e.g. maximizing the throughput. Another group also takes some users' Quality of Service (QoS) parameters into account \cite{pkt_schedule_LTE13}. Being simple and quite inefficient, efforts have been made to propose more advanced schemes \cite{QoS-resource19, resource_LTE_16}. These works, however, provide the solution only for a fixed set of system's parameters, including channel coefficients, and need to resolve the optimization problem whenever these parameters change, which incurs a huge and repetitive computation burden.\\
   As both DSS and resource allocation deal with a dynamic problem, due to changes in channel coefficients, number of users, request profiles, etc., plenty of works focused on the use of Reinforcement Learning (RL), a suitable framework to use when complex decisions need to be made on a regular basis, to solve such problems \cite{ML_Wls17, deep_wls18,DRL_survey19}.
   Early works on use of RL for resource allocation, such as \cite{simpleRL_LTE17}, have applied lookup table based RL, where a simple RL-based technique is proposed to allocate resources based on a combined metric of spectral efficiency, average delay and packet loss ratio in an LTE network. In more complex settings, lookup table approach may not be practical, as saving Q values for a big state-action space requires a very large memory, in addition to taking very long time for the agent to be trained.\\ Introducing neural networks in RL context, also known as Deep Reinforcement Learning (DRL), was a huge leap toward solving this problem \cite{deepmin_DRL15}.   
In \cite{deepRL_resource18}, DRL is applied to allocate radio and core resources  in a network slicing scenario, where a weighted sum of spectral efficiency and required Quality of Experience (QoE) by each slice is maximized. For this goal, a Long Short Term Memory (LSTM) is used to predict traffic of each slice and allocate the resources to the slices accordingly. However, since resources are allocated to slices based on the traffic prediction model, there might be some deviation from real demands, and inefficiency in resource allocation. In addition, channel states are not taken into account  in the resource allocation strategy. In \cite{DRL_hetnet}, a heterogeneous network scenario is considered where picocells and femtocells  provide area coverage cooperatively, while user  association and resource allocation is solved jointly with a DRL approach. Each user chooses among the available set of BSs and for the selected BS among the available channels. In \cite{DRL_hetnet}, QoS is defined only as a function of the minimum Signal to Interference plus Noise Ratio (SINR) constraint, without considering latency. The coexistence of WiFi and LTE-U is explored in \cite{proactive_DRL18}, where LSTM is used to predict the traffic pattern of WiFi users in the unlicensed bands. Then, some of the delay-tolerant requests in LTE are served in time-windows predicted to be occupied with lower likelihood. The LTE small Base Stations (BSs) use RL to allocate the resources such that long-term air time fairness is guaranteed for BSs and WiFi access points.\\
On the other hand, some standards allow unlicensed users to use licensed spectrum, in case they have a sharing agreement with the original license-holder, and are also equipped with Cognitive Radios (CRs), to communicate in the licensed LTE bands \cite{cognitive_share_aggregate}. 
%The approach used to implement schemes to accommodate such standards can be posed as designing CR schemes. However, unlike \cite{cognitive_share_aggregate}, we consider the problem of spectrum sharing from a different perspective.
Note that, typically, unlicensed users need to release the resources as soon as they are being used by the licensed ones. In a network with active licensed users, this can significantly increase the signaling overhead among unlicensed users or even make the resources useless when they are sparse in time and frequency.\\
% In the proposed scheme, BS of the licensed users helps unlicensed users by increasing the time of uninterrupted communication, provided that the QoS for the licensed users are guaranteed. The resource allocation scheme based on deep RL is proposed to
To resolve this challenge, in this paper we propose an alternative approach, in which the BS tries its best to help unlicensed users by leaving the longest sequence of consecutive resource blocks, in time and on the same frequency, unoccupied, provided that the required QoS for licensed users are guaranteed. This would allow an unlicensed user to use a sub-channel for communication with minimum interruption from the BS, and with minimum overhead for coordination.\\
The proposed scheme relies on DRL to  efficiently assign radio frequency resources to the requests demanded by the licensed users, while no assumption is made about any prior knowledge of the requests' distributions. In fact, the main goal of the proposed scheme is to  keep some resources unallocated (free) for the unlicensed users such that there is a higher probability that the free resources belong to the same frequency slot at successive continuous time slots. As we assume no cooperation among the licensed and the unlicensed users, such approach eases the search process of unlicensed users as they do not need to make a full sweep over the entire frequency band to find vacancies at every time slot. In addition, the overhead of handshaking signaling between the unlicensed communicating entities is reduced\footnote{Such handshakings are necessary prior to settle a communication between a sender and a receiver including  the agreement over the new frequency band.}. As RL intrinsically learns by interacting with the environment, the agent chooses a policy, acts accordingly, observes the result, and modifies its policy for the next steps. This is in contrast with the traditional approaches, which require to solve the optimization problem once in a while. Our results show that the proposed scheme is  able to maximize the average spectral efficiency, satisfy the required QoS of the received requests, and simultaneously set aside a continuous spectrum for the unlicensed users. 
% All techniques proposed for Cognitive Radio (CR) can still be used for unlicensed users in the proposed scheme. In this paper, LTE is used as a typical model for the physical and access layer. The proposed method, however, is not limited to LTE as it uses CQI table  which is available for other physical and access technologies, including 5G New Radio (NR), as well \cite{5GNR_Phy}. The proposed method can be categorized  among the non-cooperative spectrum allocation methods \cite{DSS_cognitive_survey}.
 As we do not assume any coordination among licensed and unlicensed users, the proposed method can be categorized among the non-cooperative spectrum allocation methods [1]. Furthermore, no limitation is imposed on the unlicensed users. As a result, any already-developed scheme in CR network context can be utilized by unlicensed users. It is worthwhile to mention that in this paper,  LTE is only used as a typical implementation model for the physical and access layer, and the proposed method is equally applicable to other schemes that use CQI tables, such as 5G New Radio (NR), as well \cite{5GNR_Phy}.\\
 The contributions of this paper are summarized as follows
\begin{itemize}
	\item We formulate the optimization problem for allocating time-frequency resources to the incoming requests, constrained by QoS of the users' requests in terms of latency and a notion of continuity of unallocated resources to be utilized by the unlicensed users. This can be considered as a decentralized resource sharing for the unlicensed users as we do not assume any coordination among the BS and the unlicensed users, nor do we assume any knowledge of the unlicensed users' channel states by the BS.
	\item We formulate the problem as a Markov Decision Process (MDP), where by defining the mathematical representation of the states, action and the optimization problem, we apply model-free deep RL to solve it. The requests are generated from a set of stationary, but unknown to the BS, distributions.
	\item We use experience replay and double Q networks, two famous techniques known to be efficient in deep RL, to improve the convergence of the deep RL algorithm.
	\item We assume general time-variant frequency selective channels among the BS and the User Equipments (UEs). The optimization process specifies the resource blocks associated with the requests at every step of the RL agent. This is more general than many previous papers which only determined the fraction of the allocated bandwidth for a specific user or a slice, e.g.  \cite{deepRL_resource18}.
\end{itemize}
%One of the benefits of RL is that we do not need to solve the problem many different times, as is the case for common optimization problems.
%We do not assume any knowledge of the unlicensed users' channel states or their requirements.
The rest of the paper is organized as follows. Section \ref{sec:model} describes the system model.  Mathematical formulation of the problem of interest is given in Section \ref{sec:formulation}. Section \ref{sec:DRL} provides a brief review of RL, and how it can be combined with the idea of neural networks to arrive at a deep RL scheme. The process of learning with RL is described in Section \ref{sec:RL_framework}. Numerical results along with the discussion are given in Section \ref{sec:results}. The paper is concluded in Section \ref{sec:conclusion}.
%\subsection{Notation}
%\begin{table}
%	% table caption is above the table
%	\caption{List of important parameters used throughout the paper}
%	\label{tab:notation}       % Give a unique label
%	% For LaTeX tables use
%	\begin{tabularx}{1\columnwidth}{lll}
%		\hline\noalign{\smallskip}
%		Parameter & Value& explanation (value used for simulation) \\
%		\noalign{\smallskip}\hline\noalign{\smallskip}	
%		$L$&request buffer length (fixed)& (10)\\
%		$l[n]$&request buffer length (at time n)&$0\leq l[n]\leq L$\\
%		$R$&number of total RBs given to the BS&(6)\\$\gamma$& RL discount factor&(0.99)\\
%		\noalign{\smallskip}\hline
%	\end{tabularx}
%\end{table}
\section{System Model}
\label{sec:model}
We consider a cellular network where a number of users are served by a BS at the downlink side. The channel access method is considered to be Orthogonal Frequency Division Multiple Access (OFDMA), hence each user is assigned a set of time-frequency resources during its service time. Each user's request is sent by its UE to the BS as a tuple including its required service type, with some predefined parameters, and its quantized channel estimate. Requests arrive continuously to the BS over time. We assume that the arrival process of the requests is quasi-stationary\footnote{It simply requires the requests' distributions to stay the same long enough so that the learning algorithm is able to track the changes in the distribution. As we will see in Section \ref{sec:results}, the proposed scheme converges in almost several tens of thousands of iterations, which translates into almost 10 to 20 seconds of requests' observation.} and the BS is unaware of the distribution of the arrival process.  \\
The BS is responsible for allocating time-frequency resources to the requests sent by the users. The resources consist of some time-frequency blocks called Resource Blocks (RBs). Each RB has $T$\, [s] time-width and $W$\, [Hz] frequency-width. The BS places the incoming request(s) into a buffer with fixed length $L$, as long as it has vacancy and drops the request(s) otherwise. It performs resource allocation by assigning all the RBs belonging to the current time step to some of the requests in the requests' buffer. The time steps are referred to as $\{n\}_1^N$ and  are the smallest BS time units. The requests in the buffer are indexed as $j \in \left\lbrace 1,2,\dots, L\right\rbrace $. The BS is given a total of $R$ RBs in a given time step. 
%These $R$ RBs are fixed throughout the whole allocation process. 
At each time step, we allocate the RBs to the available requests in the requests' buffer for the current time step and then proceed to a new one. Hence, we ignore the time index and refer to each of the RBs of the current time step by only the frequency index $k$, where $k \in \left\lbrace 1,2,\dots, R\right\rbrace $. Finally, we define another set of indices, called the RL steps, as follows. Consider $N$ time steps including $N\!R$ RBs. We index these RBs as $\{0,1,2,\dots, N\!R-1\}$, calling each one as an RL step, such that RL step $i$ refers to RB index $k=i\;\text{mod}\;R+1$ at time step $n=\lfloor\frac{i}{R}\rfloor+1$, where $\lfloor.\rfloor$ indicates the floor function\footnote{The name RL step refers to each step RL agent takes while interacting with the environment. The reason this name is chosen will become more clear in Section \ref{sec:RL_framework}.}.
%We also assume that at any time $n$, $K[n]$ users are sending their requests to the BS.
\subsection{Channel Model}
\label{subsec:ch_model}
In general, the channel between the BS and each user is a time-frequency varying channel. However, we assume the channel is constant in each RB. Let's denote the channel vector from the BS to the user corresponding to the $j$th request in the requests' buffer at time $n$ over $R$ RBs with $\bfh^j[n]=[h^j_1[n], h^j_2[n], \dots, h^j_R[n]]^\transp$, where $h^j_k[n] \in \bC, \forall k \in \left\lbrace 1,2,\dots, R\right\rbrace $. We assume a one-to-one correspondence between the set of users and the set of requests. In other words, every new request is issued by a user, which potentially has a new channel state. In general, there might be many requests issued by the same user. As we will see in Section \ref{sec:results}, almost all of the incoming requests are accepted by the BS. Hence, this assumption does not affect the performance of the resource allocation, since the spectral efficiency results are averaged over the entire simulation time. Taking such correspondence into account, we will use the words \textit{user} and \textit{request} interchangeably in the text.
Let's also assume that the $k$th RB is assigned to the $j$th request at time $n$. The signal received by the user corresponding to the $j$th request can then be written as
\begin{equation}
\label{eq:received_sig}
y^j[n]=h^j_k[n]x^j[n]+z^j[n],
\end{equation}
where $x^j[n]$ is the transmitted symbol and $z^j[n]$ is the additive Gaussian noise. We assume that the users experience a mixture of small-scale and large-scale fading, hence the channel coefficient $h^j_k[n]$ can be written as $h^j_k[n]=\sqrt{l^j}\zeta^j_k[n]$, where $l^j$ and $\zeta^j_k[n]$ denote the large scale and small scale fading, respectively. Large scale fading coefficient for a specific requests is assumed to stay constant as long as that request is being served and hence is not a function of $n$, while small-scale fading coefficients change after every coherence period, denoted by $\tau$\,[s]. As can be seen, for a single request $j$, small scale fading coefficient $\zeta^j_k[n]$ also depends on the RB index $k$, while large scale fading coefficient $l^j$ are the same for all RB indices. 
% We assume that the UE corresponding to the $j$th request has always access to a perfect estimate of its channel vector $\bfh^j[n]$.
We assume that the $j$th UE has always access to a perfect estimate of its channel vector $\bfh^j[n]$.
By using \eqref{eq:received_sig}, the instantaneous SINR observed by the $j$th user
%	 user corresponding to the $j$th request
	  at the $k$th RB and at time step $n$ can be written as\footnote{Without loss of generality, we assume that the interference received from the neighboring BSs are negligible. As long as there is no cooperation among the neighboring BSs, the given argument can be extended to a multicell scenario by considering interference as noise.}
\begin{equation}
\text{SINR}^j_k[n] = \frac{\bE\left| x^j[n]\right| ^2\left| h^j_k[n]\right| ^2}{\bE\left| z^j[n]\right| ^2} = \frac{P_t}{R}\frac{\left| h^j_k[n]\right| ^2}{\sigma^2_n},
\end{equation}
where $P_t$ is the BS's total transmit power, and $\sigma^2_n$ is the noise power at the UE's terminal. It is assumed that the BS is using uniform power allocation for all RBs. As  channel estimate vector $\bfh^j[n]$ is valid for a period of length $\tau$, we know that $\text{SINR}^j_k[n]$ remains constant during a coherence period.
%$\SINR^i_k[n]=\SINR^i_r[n+j], \forall j \in [1,2,\dots,\lfloor\tau/T\rfloor]$, assuming that $n$ refers to the start of a coherence period. 
We assume that each UE calculates the CQI as a function of the measured SINR, and reports it to the BS as a  $\phi$-bit integer. Each CQI corresponds to a predefined Modulation and Coding Scheme (MCS), as in LTE standard \cite{LTE_wiley09}. We assume that the CQI values are available at the BS for all the users corresponding to the requests in the BS's requests' buffer and for all the RBs at any time step\footnote{In general, depending on the patterns and the number of subcarriers given to the BS, there might be some sort of correlation among different components of the channel vector for each UE. This can be utilized by the UE to perform some sort of compression to return these values to the BS in an efficient manner.}. We denote the CQI values for the
% corresponding user of the $j$th
  $j$th user by the vector $\bfc^j[n]=[c^j_1[n], c^j_2[n], \dots, c^j_R[n]]$, where the index $n$ is used to note that this estimate is updated just at the beginning of every coherence time and hence, is in general time-dependent\footnote{More precisely, $\bfc^j\lceil n/\tau\rceil$ denotes the channel vector for each coherence period.}. Next, we define a mapping used by the BS to find the Spectral Efficiency (SE) corresponding to each CQI value.
\begin{definition}[CQI-SE Mapping]\label{def:SE mapping}
	CQI-SE mapping  $\scrT:[0, 2^{\phi-1}] \to \bR$ is defined as a fixed mapping, where $\scrT[m], \forall m \in [0, 2^{\phi-1}]$, as a function of CQI, denotes the instantaneous (per RB) achievable spectral efficiency by the user to which the RB is assigned.
\end{definition} 
Using Definition \ref{def:SE mapping}, the BS is able to calculate the number of deliverable bits for the 
%user corresponding to $j$th request
 $j$th request in the requests' buffer at RB $k$ at time $n$ as 
\begin{equation}\label{eq:num_bits}
t^j_k[n]=WT\scrT[c^j_k[n]]
\end{equation}
We assume that the number of bits calculated in \eqref{eq:num_bits} is fully deliverable to the 
%corresponding user of the $j$th request,
$j$th user.
In other words, $\scrT[c^j_k[n]]$ is below the capacity of its corresponding channel.
\subsection{Requests' Buffer}\label{sub:req_buf}
As previously mentioned, each request received by the BS includes two entries: First, its service type and second, the CQI of its corresponding user. Service type is defined as follows.
\begin{definition}[Service Types]
	A service type $m,\, m \in \left\lbrace 1,2,\dots, \theta\right\rbrace $, for some integer $\theta$, is defined by a tuple $(u^m_1,u^m_2)$, where  $u^m_1$ specifies the number of bits to be delivered for service type $m$ and $u^m_2$ refers to the maximum tolerable latency (in terms of the number of time steps) for the delivery of $u^m_1$ bits\footnote{ A typical service a user requests, such as video streaming or file download consists of many units of fixed/variable sizes that should be delivered to the user uninterruptedly. We assume that the requests for each data unit is received by the BS separately, hence the BS only observes the arrival process of the requests for such data units.}.
\end{definition}
As soon as the BS receives a single incoming request, it puts the request into the requests' buffer. The buffer's length $L$ is a fixed design parameter. As soon as the buffer becomes full, the incoming requests are dropped until at least a request is either satisfied or missed as per the following definitions. A request is said to be \textit{missed} if it has not received sufficient resources within its tolerable latency and said to be \textit{satisfied} otherwise.\\
 Let us now give a more detailed description of the requests' buffer. The $j$th element of the requests' buffer at RL step $i$, denoted by $\bfq^j[i]$, is a vector of size $R+3$. The 1st element $q^j_1[i]$ denotes the request's service type index which remains fixed as long as the $j$th request stays in the buffer. The 2nd element $q^j_2[i]$ denotes the remaining time up to the end of which the $j$th request is valid, called Time To Live (TTL) here for simplicity, it starts  from a predefined value, $u^{q^j_1[i]}_2$, and is decremented at every single time step, i.e. $q^j_2[n+1]=q^j_2[n]-1$, where $n=\lfloor\frac{i}{R}\rfloor+1$. The variable $q^j_2[n]$ is always non-negative. As soon as it becomes zero, the request is missed and removed from the buffer. The 3rd element $q^j_3[i]$ denotes the number of required bits by the $j$th request that has not been delivered so far. It starts from a predefined value, i.e. $u^{q^j_1[i]}_1$, and is decremented every time the $j$th request is given some resources. The remaining $R$ elements $\left[q^j_4[i], q^j_5[i], \dots, q^j_{R+3}[i]\right] ^\transp$ specify the number of deliverable bits by the RBs for the $j$th request, which can be written as $\bft^j[i]=\left[ q^j_4[i], q^j_5[i], \dots, q^j_{R+3}[i]\right]^\transp$ using \eqref{eq:num_bits},  and are updated at the beginning of every new coherence time. Now, we are ready to give a more formal definition of the requests' buffer:
\begin{definition}[Requests' Buffer]\label{def:req_buffer}
	The requests' buffer at RL step $i$, is defined as 
	\begin{equation}
	\clB[i]=\left\lbrace \bfq^j[i]|j \in \left\lbrace  1,2, \dots,L\right\rbrace  \right\rbrace.
	\end{equation}
\end{definition}
It should be noted that initially, the requests' buffer is empty, i.e. $\clB[0]\!\!=\!\!\left\lbrace\! \bfq^j[0]\!\!=\!\!\mathbf{0}|j\! \in\! \left\lbrace 1,\!2,\! \dots,\!L\right\rbrace\!  \right\rbrace$. As the requests arrive, the buffer is filled with the incoming requests. As we will see, the order by which a request is pushed in or popped out of the buffer depends on the resource allocation scheme and is not known in advance, hence the buffer is neither a FIFO nor a LIFO buffer.
% It is worthwhile to note that, as the BS has access to the CQI values for all the users, a vector of size $R$ for each user, the number of deliverable bits for each of the channels for all of the users can be calculated easily using the mapping $\scrT$ and the CQI returned by the UE.
 \subsection{Continuity Vector}
 As mentioned earlier, our objective is to propose a scheme which allocates the resources to the licensed users efficiently, and also keeps the continuity of the unallocated resources for the unlicensed users. First, let us define the concept of \textit{continuity vector}. 
 \begin{definition}[Continuity Vector]
 	\label{def:cont_vec}
 	The continuity vector at time step $n$ is defined as $\bfv[n]=\left[ v_1[n],v_2[n],\dots,v_R[n]\right]^\transp $, where $v_k[n]$ denotes the number of last continuous unallocated resources, from the $n$th time step backward in time and including the $n$th time step, at RB index $k$ for $k \in \left\lbrace 1,2,\dots, R\right\rbrace $.
 \end{definition}
As an example, a snapshot of the allocated and unallocated RBs for the last four time steps is given in Fig. \ref{fig:cont_vect}. The continuity vector for this figure at the last time step, i.e. $n$, is given as $\bfv[n]=[1, 1, \dots, 2]^\transp$, where $v_1[n]=1$ means, for the current time step, the first RB is unallocated, but the first RB at the previous time step is occupied. Likewise, $v_R[n]=2$ means that the $R$th RB is unallocated for the current time step and for the previous time step as well.\\
Next, we define the \textit{continuity function}
\begin{definition}[Continuity Function]
	\label{def:cont_func}
	The continuity function with respect to the parameter $C \in \mathbb{N}$ is defined as 
	\begin{equation}
	g^k_C[n]
%	=\mathbb{1}(v_k[n]\geq C)
	=\left\{\begin{array}{l}
	1 \quad \text{if} \ v_k[n]\geq C\\
	0 \quad \text{O.W.}\\
	\end{array} \right.,
	\end{equation}
	where $v_k[n]$ is the $k$th element of the continuity vector given in Definition \ref{def:cont_vec} and $C$ is a fixed integer called the continuity length.
\end{definition}
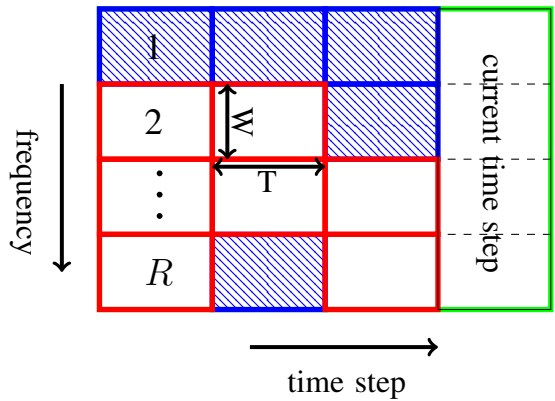
\begin{figure}
	\centering
	\usetikzlibrary{patterns}
\begin{tikzpicture}
\draw  [green][line width =2pt](-3.5,0.5) rectangle (2.5,-3.5) node (v14) {};
\draw (-2,0.5) node (v3) {} -- (-2,-3.5);
\draw (-0.5,0.5) node (v2) {} -- (-0.5,-3.5);
\draw (1,0.5) node (v1) {} -- (1,-3.5) node (v5) {};
\draw[dashed] (2.5,-0.5) -- (-3.5,-0.5) node (v4) {};
\draw [dashed](2.5,-1.5) -- (-3.5,-1.5) node (v8) {};
\draw [dashed](2.5,-2.5) -- (-3.5,-2.5) node (v12) {};
\draw [blue][pattern=north west lines, pattern color=blue,  line width=2pt] (v1) rectangle (-0.5,-0.5) node (v6) {};
\draw [blue][pattern=north west lines, pattern color=blue,  line width=2pt] (v2) rectangle (-2,-0.5) node (v7) {};
\draw [blue][pattern=north west lines, pattern color=blue,  line width=2pt] (v3) rectangle (v4);
\draw [blue][pattern=north west lines, pattern color=blue,  line width=2pt] (1,-0.5) rectangle (-0.5,-1.5) node (v9) {};
\draw [blue][pattern=north west lines, pattern color=blue,  line width=2pt] (-2,-2.5) node (v10) {} rectangle (-0.5,-3.5) node (v13) {};

\draw [red, line width=2pt] (v6) rectangle (-2,-1.5) node (v11) {};
\draw [red, line width=2pt] (v7) rectangle (v8);
\draw [red, line width=2pt] (1,-1.5) rectangle (-0.5,-2.5);
\draw [red, line width=2pt] (v9) rectangle (v10);
\draw [red, line width=2pt] (v11) rectangle (v12);
\draw [red,  line width=2pt] (v10) rectangle (-3.5,-3.5);
\draw [red,  line width=2pt] (1,-2.5) rectangle (v13);
\node[text width=3cm, rotate=270] at  (1.7, -1.6){current time step};
%\draw [->, line width=1.5 pt] (1.7, -5) -- (1.7, -3.6) {};
\draw  (v1) rectangle (v14);
\node[text width=3cm] at  (0.5, -4.5){time step};
\node[text width=1cm, rotate=270] at  (-4.5, -1.6){frequency};
\node[text width=1cm] at  (-2.4, 0){\large 1};
\node[text width=1cm] at  (-2.4, -1){\large 2};
\node[text width=1cm, rotate=270] at  (-2.7, -2.1){\LARGE \dots};
\node[text width=1cm] at  (-2.4, -3){\large $R$};
%\filldraw  (-4.5,-1.5) circle (0.1);
\draw [->, line width=1.5 pt](-1.5,-4) -- (1,-4);%time step line
\draw [->, line width=1.5pt](-4,-0.5) -- (-4,-3);%freq line

\draw [<->, line width=1.5pt](-1.8,-0.5) -- (-1.8,-1.5);%freq line (small)
\node[text width=3cm, rotate=270] at  (-1.6, -2.3){W};
\draw [<->, line width=1.5pt](-2,-1.6) -- (-0.5,-1.6);%time line (small)
\node[text width=1cm] at  (-0.9, -1.8){T};
\end{tikzpicture}
	\caption{An illustration of the continuity vector $\bfv[n]=[1, 1, \dots, 2]^\transp$ based on Definition \ref{def:cont_vec} for an example snapshot consisting of the last four time steps. The hatched and the free rectangles refer to the allocated and unallocated RBs respectively. The numbers inside rectangles refer to the RB index at each time step.}
	\label{fig:cont_vect}
\end{figure}
The continuity function is simply a step function with the value of 1 for RB index $k$ if the number of last successive unallocated resources are at least equal to $C$ and with the value of 0 otherwise.
\section{Problem Formulation}\label{sec:formulation}
\subsection{Markov Decision Process (MDP)}
Markov Decision Process (MDP) provides the mathematical tool to formalize a decision making problem with known dynamics. An MDP is defined as a tuple $\left\langle \clS, \clA, \scrP, \scrR, \gamma\right\rangle$, where $\clS$ represents the state space, $\clA$ represents the action space, $\scrP$ represents the transition probabilities, $\scrR$ represents the rewards and $\gamma$ is the discount factor \cite{sutton_RL20}.\\
In order to formulate the optimization problem in an MDP form, we need to define the \textit{state} first. The state should be defined in such a way that grasps the whole dynamics of the system, which include the adding/dropping of the requests in the buffer, the changes in the channel coefficients of the users demanding the requests, the continuity measure of the unallocated resources and the current RB index. Using the definitions given in Section \ref{sec:model}, we can define the \textit{state} as follows.
\begin{definition}[State]\label{def:state}
	The state $\bfs_i \in \clS$ is defined as
	\begin{equation}
	\bfs_i=\left[ \bfb[i]^\transp, \bfv[n]^\transp, \psi\right]^\transp,
	\end{equation}
	where $\bfb[i]$ is the requests' buffer $\clB[i]$ written in the vector form as $\bfb[i]\!\!=\!\!\left[\!{\bfq^1[i]}^\transp\!\!, {\bfq^2[i]}^\transp\!\!, \dots, {\bfq^L[i]}^\transp\right]^\transp\!\!$, $\bfv[n]$ is the continuity vector as defined before and $\psi \in \left\lbrace 1,2,\dots, R\right\rbrace $ is the  current RB index as $\psi=i\;\text{mod}\;R+1$.
\end{definition}
\begin{remark}
	The state $\bfs$, defined in Definition \ref{def:state}, has the dimension $\left( R+3\right)\times L+R+1$ which grows linearly with the buffer length $L$ and  also with the number of available resource blocks $R$.
\end{remark}
\begin{remark}
	Some parts of the state $\bfs$ defined in Definition \ref{def:state} can change only after a time step $n$, while other parts can change after every RL step $i$. The former includes $\bfv[.]$, $q_2^j[.]$, and the latter includes $\psi, q_1^j[.], q_3^j[.]$. The vector $\bft^j[.]$ can change only after several time steps corresponding to a coherence period.
\end{remark}
%\begin{definition}[Unit Vector]\label{def:unit_vec}
%	Unit vector $\bfe^{m}$ with the same dimension of the state $\bfs$ is defined as
%	\begin{equation}
%	e^{m}_{j}=\left\{\begin{array}{l}
%	1 \quad \text{if}\ j=m\\
%	0 \quad \text{O.W.}	
%	\end{array}\right.,
%	\end{equation}	
%	where $m \in \left\lbrace 1,2,\dots, \left( R+3\right)\times L+R+1\right\rbrace $.
%\end{definition}
\begin{remark}\label{rem:RB_ind}
%	We use the notation $\bfe^{m}$ in Definition \ref{def:unit_vec} to refer to a specific element of the state $\bfs$. For instance, RB index $\psi$ in Definition \ref{def:state} can be extracted from the state $\bfs$ as $\psi=\bfs^\transp\bfe^{\left( R+3\right)\times L+R+1}$. The number of delivered bits for a specific request $j$ so far, can also be written as $u_1^j-\bfs^\transp \bfe^{(j-1)(R+3)+3}$.
Based on Definition \ref{def:state}, the number of delivered bits for a specific request $j$ so far at RL step $i$, can be written as $u_1^{q_1^j[i]}-q_3^j[i]$.
\end{remark}
The next important step to define an MDP, is the \textit{action}. 
\begin{definition}[Action]\label{def:action}
	The action is defined as $a_i \in \clA=\left\lbrace 0,1,2,\dots, L\right\rbrace $, where $a_i=0$ means to leave the current RB unallocated (free), and $a_i>0$ means to allocate the current RB to the $j=a_i$th request in the requests' buffer.
\end{definition}
\begin{remark}
	For any action $a_i>0, a_i \in \clA$, taken from state $\bfs_i \in \clS$, there is a possibility that action $a_i$ refers to an empty space inside the requests' buffer. This can happen in the case where there are less requests than the value of action $a_i$ in the buffer at state $\bfs_i$. This action is called an invalid action and as a result, the corresponding RB is left unallocated.
\end{remark}
\subsection{Problem Formulation}
The next concept we need to define before arriving at the mathematical formulation for the optimization problem is the concept of \textit{spectral efficiency}. As we have already assumed, the CQI is available at the BS and hence, the spectral efficiency for each state is the number of deliverable bits in that state divided by the allocated bandwidth, as given in the following definition. 
\begin{definition}[Spectral Efficiency]\label{def:SE}
Spectral efficiency for each RB is defined as the number of deliverable bits in that RB, divided by the product of the time-width and frequency-width of a single RB, i.e. $W\!\times\! T$.
\end{definition}
As we will see later in Section \ref{sec:RL_framework}, the resource allocation procedure is such that a state-action pair $(\bfs_i,a_i)$ specifies the assignment of a single RB to a specific request. This notion of spectral efficiency can be calculated using the following remark.
\begin{remark}[Spectral Efficiency]\label{th:SE}
	Spectral efficiency, as a function of the state $\bfs_i$ and the action $a_i$ can be calculated as
%	\begin{equation}\label{eq:SE}
%	\text{SE}(\bfs,a)=\left\{\begin{array}{cc}
%	\frac{\bfs^\transp\bfe^{\left( a-1\right) \left( R+3\right) +\psi+3}}{WT}& \text{if}\ \ a>0\\
%	0&\text{if}\ \ a=0
%	\end{array}\right.,
%	\end{equation}
\begin{equation}\label{eq:SE}
\text{SE}(\bfs_i,a_i)=\left\{\begin{array}{cc}
\frac{q^{a_i}_{\psi+3}[i]}{WT}& \text{if}\ \ a_i>0\\
0&\text{if}\ \ a_i=0
\end{array}\right.,
\end{equation}
	where $\psi$ and $q^{a_i}_{\psi+3}[i]$ are both parts of the state $\bfs$ according to Definition \ref{def:state}.\\
	For the case where $a_i=0$, no request is selected and the RB is left free, hence we should have $\text{SE}=0$. This can be seen from \eqref{eq:SE} as well.\\
		For any other case, i.e. $a_i>0$, a request is chosen from the requests' buffer by the action $a_i$ as the buffer index. The selected request should be $\bfq^{a_i}[i]$ according to Definition \ref{def:req_buffer}. The vector $\bft^{a_i}[i]$ which refers to the last $R$ components of $\bfq^{a_i}[i]$ contains the number of deliverable bits for every RB for the selected request. Finally, the current RB index $\psi$ corresponds to the $\psi+3$th component of the request $\bfq^{a_i}[i]$. Based on the discussion given in Section \ref{sub:req_buf}, this component can be written as $q^{a_i}_{\psi+3}[i]$.
	%	Based on Definition \ref{def:state},  index $\left( a-1\right) \left( R+3\right)$ refers to the 1st element of $\bfq^a[i]$ in $\bfs$.
%		 Next, the number of deliverable bits for the chosen request and for the current RB should be selected. The vector $\bft^a[i]$ which refers to the last $R$ components of $\bfq^a[i]$ contains the number of deliverable bits for every RB for the selected request. The current RB index $\psi$ corresponds to the $\psi+3$th component of the request $\bfq^a[i]$. Based on the discussion in Section \ref{sub:req_buf}, this component can be written as $q^a_{\psi+3}[i]$.
%	%	 $\bfs^\transp\bfe^{\left( a-1\right) \left( R+3\right) +\psi+3}$.
%		  This completes the proof.
\end{remark}
\begin{remark}
	The notion of Spectral Efficiency defined in \eqref{th:SE} is in fact an optimistic metric. This is the case since the requests to which resources are allocated currently might be missed in the future. This should be taken into account and will be handled shortly.
\end{remark}
%\begin{proof}
%	Case1: $a=0$. In this case, no request is selected and the RB is left free, hence we should have $\text{SE}=0$. This can be seen from \eqref{eq:SE} as well.\\
%	Case2: $a>0$. In this case, a request is chosen from the requests' buffer by the action $a$ as the buffer index. The selected request should be $\bfq^a[i]$ according to Definition \ref{def:req_buffer}. 
%%	Based on Definition \ref{def:state},  index $\left( a-1\right) \left( R+3\right)$ refers to the 1st element of $\bfq^a[i]$ in $\bfs$.
%	 Next, the number of deliverable bits for the chosen request and for the current RB should be selected. The vector $\bft^a[i]$ which refers to the last $R$ components of $\bfq^a[i]$ contains the number of deliverable bits for every RB for the selected request. The current RB index $\psi$ corresponds to the $\psi+3$th component of the request $\bfq^a[i]$. Based on the discussion in Section \ref{sub:req_buf}, this component can be written as $q^a_{\psi+3}[i]$.
%%	 $\bfs^\transp\bfe^{\left( a-1\right) \left( R+3\right) +\psi+3}$.
%	  This completes the proof.
%\end{proof}
%Next, we are going to define a simple function that holds the value of PDU length for each service type.
%\begin{definition}[PDU Size]
%	We define the function $h[i]$, where $i \in \clT$ as a fixed and known mapping from the service type space $\clT$. 
%\end{definition}
Now, we formulate an optimization problem which aims to simultaneously maximize the average spectral efficiency, for the delivered requests, keep some RBs unallocated for unlicensed users in a continuous manner as defined in Definition \ref{def:cont_func}, and minimize the number of missed requests. First, we need to quantify the sum of spectral efficiencies for a single time step $n$. This can be written as 
\begin{equation}\label{eq:obj_1}
\sum_{\lfloor\frac{i}{R}\rfloor+1=n}\text{SE}(\bfs_i,a_i)-\frac{\zeta[n]}{WT},
\end{equation} 
where $\zeta[n]$ denotes the number of allocated bits corresponding to the requests that are missed at time $n$. The first term in \eqref{eq:obj_1} is simply a sum over the spectral efficiencies for different RBs belonging to the $n$th time step based on Remark \ref{th:SE} and the second term refers to the share of missed requests at time $n$ that has previously been taken into account in the spectral efficiency of previous time steps and must be deducted from it. The variable $\zeta[n]$ can be written as $\zeta[n] = \sum_{m \in \clZ(\tilde{\bfs})}\left( u_1^{q_1^m[i]}-q_3^m[i]\right) $ where $\tilde{\bfs}$ can be any state from the set $\tilde{\bfs}=\left\lbrace \bfs_i|\lfloor\frac{i}{R}\rfloor+1=n\right\rbrace$\footnote{This is due to the fact that as soon as some requests are missed at time step $n$, they are known as missed requests at every state in that time step.} and $\clZ(\tilde{\bfs})$ is the set of all missed requests in the requests' buffer at state $\tilde{\bfs}$. Next, we take into account the effect of continuity for the $n$th time step. Using Definition \ref{def:cont_func}, this can be written as
\begin{equation}\label{eq:obj_2}
\sum_{k=1}^Rg^k_C[n].
\end{equation}
Finally, we need to take into account the effect of latency. This can be written using the concept of TTL defined earlier. We set this term such that its value approaches zero in the case where the request with the minimum latency is about to be missed, i.e. its TTL is close to zero. This can be written as 
\begin{equation}\label{eq:obj_3}
1- \exp\left(  -\delta\min_{l\in \clY[n]}\left( \frac{q_2^l[n]}{u_2^{q_1^l[n]}}\right) \right),
\end{equation}
where $\clY[n]$ refers to the set of nonempty requests in the requests' buffer at time $n$ and $\delta$ is a parameter which controls how fast or how slow \eqref{eq:obj_3} moves towards zero. The TTL values are normalized to the maximum possible TTL value of each service type to make sure that the scheme is fair across different service types.
%It can be seen that as soon as the normalized minimum TTL value for all the requests in $\clY[n]$ approaches zero, \eqref{eq:obj_3} approaches zero as well.
Now, we are ready to combine \eqref{eq:obj_1}, \eqref{eq:obj_2} and \eqref{eq:obj_3} to formulate the optimization problem as follows
\begin{align}\label{eq:objective}
&\max_{\left\lbrace a_i\right\rbrace _{i=0}^{NR-1}} \limsup\limits_{N\to \infty}\frac{1}{N}\sum_{n=1}^{N}\left\lbrace  \left[ \alpha\!\!\!\sum_{\lfloor\frac{i}{R}\rfloor+1=n}\!\!\left(\!  \text{SE}(\bfs_i,a_i)-\frac{1}{RWT}\sum_{m\in \clZ(\bfs_i)} \left( u_1^{q_1^m[i]}-q_3^m[i]\right)\!\right)  +\beta\sum_{k=1}^Rg^k_C[n]\right]\right.\nonumber\\
&\times\left. \left[ 1- \exp\left(  -\delta\min_{l\in \clY[n]}\left( \frac{q_2^l[n]}{u_2^{q_1^l[n]}}\right) \right) \right] \right\rbrace,
\end{align}
where $\limsup(.)$ refers to the supremum limit. As the objective function in \eqref{eq:objective} is always non-negative and bounded, $\limsup(.)$ is used to make sure potential oscillating solutions to \eqref{eq:objective} are also included. The objective function in \eqref{eq:objective} consists of a sum, over $N$ time steps, of three terms where the third one is multiplied by the linear combination of the other two. As the minimum normalized TTL value approaches $0$, the third term goes to zero, based on the value of the parameter $\delta$, resulting in the share of the entire $n$th term to be $0$. This serves as a maximum tolerable latency constraint.\footnote{We have tested several other forms for the objective function including the form of three additive terms. Finally, it became clear that the case with the best answers in RL formulation was the one given here, i.e. a linear combination of the first and second terms multiplied by the third term where the third term is in an exponential form.}
The three terms in \eqref{eq:objective} all depend on the sequence of actions $\left\lbrace a_i\right\rbrace _{i=0}^{NR-1}$. Solving \eqref{eq:objective} at one shot is not possible, since the allocation needs to be done online. The solution should be provided for every time step $n$ as time goes on. Even for a limited value of $N>1$, there is no causal one shot solution to \eqref{eq:objective}. Apart from the causality issue, the objective function in \eqref{eq:objective} consists of a Nonlinear Integer Programming. Also, as $N$ grows larger, the dimension of the problem grows as large. Thus, this problem cannot be solved using common optimization frameworks. The online solution, however, can be found by formulating the problem into an MDP form and solving it using RL, which will be discussed in detail in the next two sections.
\section{Deep Reinforcement Learning}
\label{sec:DRL}
\subsection{Introduction to RL}
 As stated earlier in Section \ref{sec:formulation}, MDP formalizes a decision making problem with known dynamics. Based on Definitions \ref{def:state} and \ref{def:action}, it can be seen that as soon as any action $a_i \in \clA$ is taken from the state $\bfs_i \in \clS$, the state changes to $\bfs_{i+1}$, since any new action results in a change in $\psi$ and hence $\bfs_i$. The state also changes at the beginning of every new time step due to the reduction in the requests' TTL values and at every new coherence time due to change in CQI values. It also changes as soon as a new request arrives, provided that the buffer has vacancy.  In summary, the state $\bfs_i$ is seen to be a sufficient statistic for the next state $\bfs_{i+1}$. So, the problem of resource allocation in \eqref{eq:objective} can be modeled using an MDP. We are not going to calculate the probabilities for the mentioned transitions of the states, as we will exploit model-free RL which relies on samples and not on the model.\\
 The agent, which interacts with the environments, at every step $i \geq 1$ starts from a state $s_i$ and takes an action $a_i$, receives an immediate reward $r_i$ and lands in a new state $s_{i+1}$ \cite{szepesvari2010algorithms}. As mentioned in Section \ref{sec:model}, we differentiate between the time step referring to the time each RB spans, which was previously indexed by $n$, and the RL step which is indexed by $i$. More clearly, each time step consists of exactly $R$, RL steps. This difference is important, since the resource allocation, as will be explained in Section \ref{sec:RL_framework}, is done based on the latter  .\\
Before moving on, let us briefly explain the concept of \textit{episode}. Episodes are the subsequences that the agent-environment interaction is broken into. Each episode ends in a state called the \textit{terminal state} \cite{sutton_RL20}.  A very well-known example of episodic environment is observed in classic games such as chess where an episode terminates as soon as the game is either won or lost . As we will see in Section \ref{sec:results}, the RL agent is trained in an episodic setting.
%Although our environment is naturally a continuing one, in contrast to an episodic one, we will see in Section \ref{sec:results} that breaking it into several episodes can help training the RL agent. \Red{do not forget to explain this in results chapter}
\subsection{Q Learning with Function Approximation}
In order to make proper decisions, the agent needs to have a criterion for the goodness of each state-action pair which is quantified using \textit{action-value function}. Typically, the value function for state-action pair $(s_i,a_i)$ is denoted by $Q(s_i,a_i$) stored in a table. In complex settings where tabular approach cannot hold all state-action value pairs, due to large memory requirement or very long exploration time, neural networks can be used to represent state-action pairs. Such approach is well supported by the fact that neural networks are universal function approximators \cite{neural_net_universal01, neural_net_express17}. Q function in this case is written as $Q(s_i, a_i, \bfw)$, where $\bfw$ denotes the weights of the neural network. In this case, weights at step $i$ are updated such that the following mean-squared error is minimized: 
\begin{equation}\label{eq:loss}
L_i(\bfw_i)=\bE_{s_i, a_i, r_i}\left[\left(  \bE_{s_{i+1}}[y|s_i,a_i]-Q(s_i, a_i, \bfw_i)\right) ^2\right] ,
\end{equation}
where $y_i=r_i+\gamma \max_{a}Q(s_{i+1},a,\bfw^-_i)$ is the target, $\bfw_i$ refers to the weights at iteration $i$ and $\bfw^-_i$ refers to the weights at some previous iteration. The well known Q learning weights update rule can be derived by differentiating \eqref{eq:loss} with respect to the weights $\bfw_i$, replacing the expectation with single samples and updating the weights every step, $\bfw^-_i=\bfw_{i-1}$, as follows \cite{deepmind_RL15}
\begin{equation}
\label{eq:w_update}
\bfw_{i}= \bfw_{i-1}+\kappa\left( r_i+\gamma\max_{a}Q(s_{i+1},a,\bfw_{i-1})-Q(s_i,a_i,\bfw_i)\right) \nabla_{\bfw_i} Q(s_i,a_i,\bfw_i),
\end{equation}
where $\kappa$ is the learning rate and $\nabla_{\bfw_i}(.)$ denotes differentiation with respect to $\bfw_i$. This simple rule is data inefficient, since we use each state-action pairs only once, and more importantly, might result in instability. We will shortly review two famous methods that are used to stabilize the updates and improve the convergence behavior of deep RL.
\subsection{Experience Replay}
\label{sub:replay}
Inspired by \cite{deepmind_RL15}, sequences of the observed state, action, reward and next state are stored in a memory called the \textit{replay memory} as $\left\lbrace \bfs_i, a_i, r_i, ,\bfs_{i+1}, \nu_i\right\rbrace$, where $\nu_i$ is a binary variable indicating whether the next state is a terminal state, $\nu_i = \text{True}$  or not, $\nu_i=\text{False}$. During training, minibatch of size $M$ is taken randomly from the replay memory every time the network is trained. This has the benefit of decorrelating the sequences of observations and actions. In other words, the network is trained with states and actions belonging to different points on the time line and thus, improve the convergence behavior of the neural network.
\subsection{Target Network}
Another useful method to stabilize neural network in RL is to use a separate network, called the \textit{target network}, for generating the targets for the weights update in \eqref{eq:w_update}. The second network's weights are updated less frequently, at a frequency of $N_t$ in terms of the number of RL steps, compared to the main network. During the interval of each two successive updates of the main network, the target network's weights are kept fixed \cite{deepmind_RL15}. So, the target in this case will be
\begin{equation}\label{eq:target}
y_i=r_i+\gamma \max_{a}\hat{Q}(s_{i+1},a,\bfw^-_i)
\end{equation}
%\begin{equation}
%\label{eq:Q_update_target_old}
%Q_{\text{new}}(\bfs_j,a_j,\bfw) = r_j + \gamma\max_{a^\prime}\hat{Q}(\bfs_{j+1},a^\prime_j,\bf\hat{w}),
%\end{equation}
%\begin{equation}
%\label{eq:w_update_target}
%\bfw^+= \bfw^-+\kappa\left( r_i+\gamma\max_{a}\hat{Q}(s_{i+1},a,\hat{\bfw})-Q(s_i,a_i,\bfw^-)\right) \nabla_{\bfw^-} Q(s_i,a_i,\bfw^-).
%\end{equation}
where $\bfw^-_i$ and $\hat{Q}(s_{i+1},a,\bfw^-_i)$ refer to the weights and the output of the target network, respectively.\\
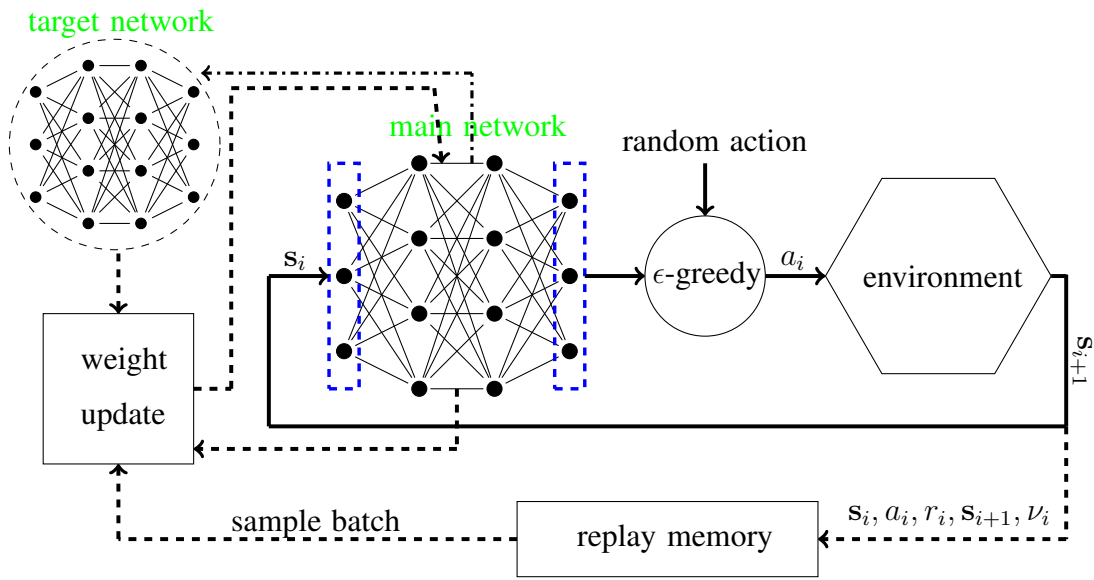
\begin{figure}
	\centering
	\begin{tikzpicture}
\def\R{1}
\def \xtargetnet {-4.4}
\def \ytargetnet {2.5}
\def \xenv {6.9}
\def \yenv {2.5}

\pgfarrowsdeclare{arcs}{arcs}{...}
{
	\pgfsetdash{}{0pt} % do not dash
	\pgfsetroundjoin   % fix join
	\pgfsetroundcap    % fix cap
	\pgfpathmoveto{\pgfpoint{-10pt}{10pt}}
	\pgfpatharc{180}{260}{10pt}
	\pgfpatharc{90}{180}{10pt}
	\pgfusepathqstroke
}
%%%% neural net (main)
\begin{scope}[xshift=0 cm]
\node[text width=3cm, color=green] at  (1.1, 4.5){main network};

%% input
%% nodes
\node (A) at (-1, 3.5) {};
\node (B) at (-1, 2.5) {};
\node (C) at (-1, 1.5) {};
%% circles
\filldraw (A) circle (\R mm);
\filldraw (B) circle (\R mm);
\filldraw (C) circle (\R mm);

%% 1st layer
%% nodes
\node (D) at (0, 4) {};
\node (E) at (0, 3) {};
\node (F) at (0, 2) {};
\node (G) at (0, 1) {};
%% circles
\filldraw (D) circle (\R mm);
\filldraw (E) circle (\R mm);
\filldraw (F) circle (\R mm);
\filldraw (G) circle (\R mm);
%% connections
\draw [-] (A) edge (D) ;
\draw [-] (A) edge (E) ;
\draw [-] (A) edge (F) ;
\draw [-] (A) edge (G) ;

\draw [-] (B) edge (D) ;
\draw [-] (B) edge (E) ;
\draw [-] (B) edge (F) ;
\draw [-] (B) edge (G) ;

\draw [-] (C) edge (D) ;
\draw [-] (C) edge (E) ;
\draw [-] (C) edge (F) ;
\draw [-] (C) edge (G) ;

%% 2nd layer

%% nodes
\node (H) at (1, 4) {};
\node (I) at (1, 3) {};
\node (J) at (1, 2) {};
\node (K) at (1, 1) {};
%%circles
\filldraw (H) circle (\R mm);
\filldraw (I) circle (\R mm);
\filldraw (J) circle (\R mm);
\filldraw (K) circle (\R mm);
%% connections
\draw [-] (D) edge (H) ;
\draw [-] (D) edge (I) ;
\draw [-] (D) edge (J) ;
\draw [-] (D) edge (K) ;

\draw [-] (E) edge (H) ;
\draw [-] (E) edge (I) ;
\draw [-] (E) edge (J) ;
\draw [-] (E) edge (K) ;

\draw [-] (F) edge (H) ;
\draw [-] (F) edge (I) ;
\draw [-] (F) edge (J) ;
\draw [-] (F) edge (K) ;

\draw [-] (G) edge (H) ;
\draw [-] (G) edge (I) ;
\draw [-] (G) edge (J) ;
\draw [-] (G) edge (K) ;

%% output
%% nodes
\node (L) at (2, 3.5) {};
\node (M) at (2, 2.5) {};
\node (N) at (2, 1.5) {};
%% circles
\filldraw (L) circle (\R mm);
\filldraw (M) circle (\R mm);
\filldraw (N) circle (\R mm);
%% connections
\draw [-] (H) edge (L) ;
\draw [-] (H) edge (M) ;
\draw [-] (H) edge (N) ;

\draw [-] (I) edge (L) ;
\draw [-] (I) edge (M) ;
\draw [-] (I) edge (N) ;

\draw [-] (J) edge (L) ;
\draw [-] (J) edge (M) ;
\draw [-] (J) edge (N) ;

\draw [-] (K) edge (L) ;
\draw [-] (K) edge (M) ;
\draw [-] (K) edge (N) ;
%%%%%%%%%%%%%%%%%%%%%
%input box
\draw(-1.2, 4) [blue, dashed, line width=1.3 pt] rectangle (-0.8, 1);
%outut box
\draw(1.8, 4) [blue, dashed, line width=1.3pt] rectangle (2.2, 1);
%\node (O) at (2, 2.5) {};
%\node (P) at (4, 2.5) {};
%\draw [->] (O) edge (P){} ;
%\node[text width=2cm] at  (3.4,2.3){$\argmax$};
%%% epsilon greedy circle
\draw (3.8, 2.5) circle (8 mm);
\node[text width=2cm] at  (4.1, 2.5){$\epsilon$-greedy};
%\draw [->] (2.2, 2.5) edge (3, 2.5) {};
%\draw[-arcs,line width=1pt] (2.2, 2.5)  -- (3, 2.5);
\draw[->, line width=1.5 pt](2.2, 2.5)  -- (3, 2.5);%neural net to epsilon
\node[text width=3cm] at  (4.2, 4.3){random action};
\draw [->, line width=1.5 pt] (3.8, 4) -- (3.8, 3.3) {};%random action to epsilon
\draw [->, line width=1.5 pt] (4.6, 2.5) -- (5.4, 2.5) {};%epsilon to environment
\end{scope}
%%% a_i
\node[text width=1cm] at  (5.3, 2.7){$a_i$};
%%%%%%%%% environment

\begin{scope}[xshift=\xenv cm, yshift=\yenv cm]
%\draw (4, 3.5)
\newdimen\RR
\RR=1.5cm
\draw (0:\RR) \foreach \x in {60,120,...,360} {  -- (\x:\RR) };
\foreach \x/\l/\p in
{ 60/{}/above,
	120/{}/above,
	180/{}/left,
	240/{}/below,
	300/{}/below,
	360/{}/right
};
%\node[inner sep=1pt,circle,draw,fill,label={\p:\l}] at (\x:\RR) {};
\node[text width=2cm] at  (0, 0){environment};

\end{scope}
%% env net connection
\draw [-, line width=1.5 pt] (8.4, 2.5) edge (8.6, 2.5) {};
\draw [-, line width=1.5 pt] (8.6, 2.52) edge (8.6, 0.5) {};

\draw [-, line width=1.5 pt] (8.6, 0.5) -- (-2, 0.5) {};
\draw [-, line width=1.5 pt] (-2, 0.5) -- (-2, 2.5) {};
\draw [->, , line width=1.5 pt] (-2, 2.5) -- (-1.2, 2.5) {};
%%s_{i+1}
\node[text width=2cm, rotate= 270] at  (8.8, 0.7){$\bfs_{i+1}$};
%%% s_i
\node[text width=2cm] at  (-0.8, 2.7){$\bfs_i$};
%%%% 2nd connection to replay buffer
%%incoming to buffer
\draw [-, line width=1.5 pt, dashed] (8.6, 0.5)-- (8.6, -1) {};
\draw [->, line width=1.5 pt, dashed] (8.6, -1) -- (5.3, -1) {};
%%outgoing from buffer
\draw [-, line width=1.5 pt, dashed] (1.3, -1) -- (-4, -1) {};
\draw [->, line width=1.5 pt, dashed] (-4, -1) -- (-4, 0) {};

\node[text width=2cm] at  (6.7, -0.7){$\bfs_i, a_i, r_i, \bfs_{i+1}, \nu_i $};
%% sample batch
\node[text width=3cm] at  (-1, -0.8){sample batch};
%%%%%%%%%%% weight update
\draw (-5, 0)  rectangle (-3, 2);
\node[text width=2cm] at  (-3.5, 1){weight update};
\draw [-, line width=1.5 pt, dashed] (-3, 1)-- (-2.5, 1);
\draw [-, line width=1.5 pt, dashed] (-2.5, 1)-- (-2.5, 5);
\draw [-, line width=1.5 pt, dashed] (-2.5, 5)-- (0.2, 5);
\draw [->, line width=1.5 pt, dashed] (0.2, 5)-- (0.3, 4);

%%%%%%%%%%%% replay buffer
\draw (1.3, -1.5)  rectangle (5.3, -0.5);
\node[text width=3cm] at  (3.6, -1){replay memory};
%% target net to weight update connection
\draw [->, line width=1.5 pt, dashed] (-4, 2.8) -- (-4, 2) {};
%% main net to weight update connection
\draw [-, line width=1.5 pt, dashed] (0.5, 1) -- (0.5, 0.2) {};
\draw [->, line width=1.5 pt, dashed] (0.5, 0.2) -- (-3, 0.2) {};
%% main network to target net connection (for weight updates of the target net)
\draw [-, line width=1.2 pt, dashdotted] (0.7, 4) -- (0.7, 5.2) {};
\draw [->, line width=1.2 pt, dashdotted] (0.7, 5.2) -- (-2.9, 5.2) {};
%%%% neural net (target)
\begin{scope}[xshift=\xtargetnet cm, yshift=\ytargetnet cm, scale= 0.7]
\draw (0.5, 2.5) [dashed] circle (20 mm);
\node[text width=3cm, color=green] at  (1, 4.8){target network};

%% input
%% nodes
\node (A) at (-1, 3.5) {};
\node (B) at (-1, 2.5) {};
\node (C) at (-1, 1.5) {};
%% circles
\filldraw (A) circle (\R mm);
\filldraw (B) circle (\R mm);
\filldraw (C) circle (\R mm);

%% 1st layer
%% nodes
\node (D) at (0, 4) {};
\node (E) at (0, 3) {};
\node (F) at (0, 2) {};
\node (G) at (0, 1) {};
%% circles
\filldraw (D) circle (\R mm);
\filldraw (E) circle (\R mm);
\filldraw (F) circle (\R mm);
\filldraw (G) circle (\R mm);
%% connections
\draw [-] (A) edge (D) ;
\draw [-] (A) edge (E) ;
\draw [-] (A) edge (F) ;
\draw [-] (A) edge (G) ;

\draw [-] (B) edge (D) ;
\draw [-] (B) edge (E) ;
\draw [-] (B) edge (F) ;
\draw [-] (B) edge (G) ;

\draw [-] (C) edge (D) ;
\draw [-] (C) edge (E) ;
\draw [-] (C) edge (F) ;
\draw [-] (C) edge (G) ;

%% 2nd layer

%% nodes
\node (H) at (1, 4) {};
\node (I) at (1, 3) {};
\node (J) at (1, 2) {};
\node (K) at (1, 1) {};
%%circles
\filldraw (H) circle (\R mm);
\filldraw (I) circle (\R mm);
\filldraw (J) circle (\R mm);
\filldraw (K) circle (\R mm);
%% connections
\draw [-] (D) edge (H) ;
\draw [-] (D) edge (I) ;
\draw [-] (D) edge (J) ;
\draw [-] (D) edge (K) ;

\draw [-] (E) edge (H) ;
\draw [-] (E) edge (I) ;
\draw [-] (E) edge (J) ;
\draw [-] (E) edge (K) ;

\draw [-] (F) edge (H) ;
\draw [-] (F) edge (I) ;
\draw [-] (F) edge (J) ;
\draw [-] (F) edge (K) ;

\draw [-] (G) edge (H) ;
\draw [-] (G) edge (I) ;
\draw [-] (G) edge (J) ;
\draw [-] (G) edge (K) ;

%% output

%% output
%% nodes
\node (L) at (2, 3.5) {};
\node (M) at (2, 2.5) {};
\node (N) at (2, 1.5) {};
%% circles
\filldraw (L) circle (\R mm);
\filldraw (M) circle (\R mm);
\filldraw (N) circle (\R mm);
%% connections
\draw [-] (H) edge (L) ;
\draw [-] (H) edge (M) ;
\draw [-] (H) edge (N) ;

\draw [-] (I) edge (L) ;
\draw [-] (I) edge (M) ;
\draw [-] (I) edge (N) ;

\draw [-] (J) edge (L) ;
\draw [-] (J) edge (M) ;
\draw [-] (J) edge (N) ;

\draw [-] (K) edge (L) ;
\draw [-] (K) edge (M) ;
\draw [-] (K) edge (N) ;
\end{scope}
%%%% dots (time direction)
%
%\filldraw (3,1.5) circle (0.5 mm);
%\filldraw (3.2,1.5) circle (0.5 mm);
%\filldraw (3.4,1.5) circle (0.5 mm);
%
%%% central unit
%\draw (0, 0)  rectangle (5, 4);
%%%% horizontal lines
%\draw [line width=0.5mm] (0, 3)--(5, 3);%line 1
%\draw [line width=0.5mm] (0, 2)--(5, 2);%line 2
%\draw [line width=0.5mm] (0, 1)--(5, 1);%line 1
%%%%vertical lines
%\draw [line width=0.5mm] (1, 4)--(1, 0);%line 1
%\draw [line width=0.5mm] (2, 4)--(2, 0);%line 1
%%%% horizontal description
%\node[text width=3cm, align = center] at  (2.5,-1){time slot};

%%%%vertical description
%\node [text width=3cm, rotate = 270] at (-1, 1.5) {frequency slot};
%\node (C) at (-1/2, 3) {};
%\node (D) at (-1/2, 1) {};
%\draw [->] (C) edge (D) ;
%
%%%%%%%%% RB indices
%\node[text width=5cm, align = center, color = red] at  (1/2,3.5){1};
%\node[text width=5cm, align = center, color=red] at  (1/2,2.5){2};
%\node[text width=5cm, align = center, color=red] at  (1/2,0.5){R};
%%%%%% snake-like arrow
%\node (E) at (1/2+0.2, 3.5){};
%\node (F) at (1/2+0.2, 0.5){};
%\node (G) at (3/2+0.2, 3.5){};
%\node (H) at (3/2+0.2, 0.5){};
%\node (I) at (5/2+0.2, 3.5){};
%\node (J) at (5/2+0.2, 0.5){};
%\draw [line width = 0.9, color = blue][->] (E) edge (F);
%\draw [line width = 0.9, color = blue][->] (F) edge (G);
%\draw [line width = 0.9, color = blue][->] (G) edge (H);
%\draw [line width = 0.9, color = blue][->] (H) edge (I);
%\draw [line width = 0.9, color = blue][->] (I) edge (J);

\end{tikzpicture}
	% 	\centering
	\caption{An illustration of the deep RL mechanism using experience replay and target network to make the network more stable}
	\label{fig:deepRL}
\end{figure}
We finalize this section by discussing the strategy used to take the actions.
\subsection{$\epsilon$-greedy Strategy}
\label{sub:epsilon_greedy}
A known issue in RL is \textit{exploration-exploitation dilemma}, which can be simply explained as follows. How should we make a balance between exploring new actions from the known states, which might end up in new unexplored states, and exploiting the best known actions. \textit{$\epsilon$-greedy} strategy is a common strategy which targets this dilemma. It simply starts from a random action selection strategy and gradually decreases the randomness and instead, increases the chance of selecting the best known action for each state.\\
In this paper, the two ideas that have just been explained, namely the Experience Replay and the Target Network, along with the $\epsilon$-greedy strategy are used. The detailed algorithm is given in Alg. \ref{alg:main}. Also an illustration of the deep RL mechanism is depicted in Fig. \ref{fig:deepRL}. The only remaining issue will be the rewarding mechanism which is discussed in the next section.
%In the next two sections, we shortly cover the two important methods that are commonly used in today's deep RL methods. These methods help stabilizing the neural net and improve deep RL convergence.
\section{RL Framework for Resource Allocation}
\label{sec:RL_framework}
In this section, the remaining elements that let one solve \eqref{eq:objective} in an RL framework are given. The resource allocation scheme is as follows. The agent moves along the two dimensions of frequency and time and allocates the resources in a repetitive procedure. At RL step $i$, the agent is given an action $a_i$, which is the output of the  $\epsilon$-greedy strategy. If the action is a zero action, i.e. $a_i=0$, the $k=i\;\text{mod}\;R+1$th RB is left unallocated. Otherwise, the action is used as an index and the $k$th RB is allocated to the $j=a_i$th request. There is a special case where the action is invalid, meaning that $a_i$ refers to an empty space in the requests' buffer. In this case, the RB is left unallocated as well. The difference with the $a_i=0$ case is then about the reward given to the agent in these two cases. Once the action is taken by the agent, the state $\bfs$ is updated and fed back to the neural network as input. This process goes on as long as the resource allocation is required.
% The allocation process can be seen in Fig. \ref{fig:time_freq_table}.
\subsection{Rewarding Mechanism}
What really matters in directing the RL agent towards obtaining the objective in \eqref{eq:objective} is the rewarding mechanism which is given in Alg.\,\ref{alg:rewards}. Except for the special cases, i.e. empty buffer and invalid action, the reward consists of three terms in accordance with \eqref{eq:objective}. There are, however, some differences with \eqref{eq:objective}. The first one is the removal of the share in spectral efficiency due to missed requests in the rewarding mechanism. Its role, however, is still being played by the third term in \eqref{eq:objective} which tries to keep missed requests as low as possible. In case a request is missed, the removed part of the first term approaches $0$, and in fact they both act in the same way. The second difference is the normalization of $\text{SE}(\bfs,a)$ to its maximum possible value in Alg.\,\ref{alg:rewards} in calculating $r^1$. This lets $\alpha, \beta$ have a more balanced effect and hence, facilitates their fine adjustment. Moreover, we accumulate the rewards corresponding to the first and the second terms for different RBs at each time step. At the end of the time step, the accumulated first and second terms form a weighted sum and the result is multiplied by the third term which is given as the reward to the agent. The reason for delaying the reward up to the end of the time step is that the first and the second terms cannot be nonzero simultaneously. This can lead the agent towards a dominating policy, where only one of the first or the second terms dominates.\\
We finish this section with a small note.
As previously noted in Section \ref{sec:DRL}, each time step consists of $R$ RL steps.
The framework could have been designed in a way that all RBs at a time step are allocated at once. This, however, would require a neural network with the output size of $R$ times as large which takes many more iterations to be trained\footnote{This is in case one-hot encoding is used as the output coding for the neural network.}. Based on the rewarding mechanism described in this section and the deep RL mechanism described in Section \ref{sec:DRL}, we are ready to describe the resource allocation scheme detailed in Alg.\,\ref{alg:main}.\\

%\begin{figure}[b]
%	\centering
%	%	\includegraphics[width=\textwidth,height=6cm]{Fig4.eps}
%	\input{time_freq_table.tex}
%	% 	\centering
%	\caption{Time-frequency resources that are allocated by the RL agent. The arrows show the order of the steps taken by the agent to allocate the resources to the requests in the buffer.
%		the numbers show the RB indices $\left\lbrace 1,2, \dots, R\right\rbrace $ for the current time step. Each time step consists of exactly $R$, RL steps.}
%	\label{fig:time_freq_table}
%\end{figure}
% \subsection{Rewards}
% \label{sub:rewards} 
\begin{algorithm}[t]
	\caption{Rewarding mechanism for the proposed deep RL Algorithm (at time step $n\!=\!\lfloor \frac{i}{R}\rfloor\!+\!1$)}
	\label{alg:rewards} 
	\begin{algorithmic}[1]
		\State Input: $\bfs_i, a_i$, Output: $r_i$; $k=i\;\text{mod}\;R+1 \in \left\lbrace 1,2,...,R\right\rbrace$
		\State Initialize $r^1=r^2=r^3 = 0$
		\For{$k \in \left\lbrace 1,2,...,R\right\rbrace $}
			\If {the requests' buffer is empty}
				\State return $0$
			\Else 
				\If {$a_i$ is an invalid action}
					\State return $-1$
%				\ElsIf{output of the neural network is $\bf0$}
%					\State return $-1$
				\Else
					\State $r ^1 += \text{SE}(\bfs_i,a_i)/\text{SE}_{\text{max}}, r^2 +=g^k_C[n]$
					\State return 0
				\EndIf
			\EndIf
		\EndFor
		\State $r^3=\left[ 1- \exp\left(  -\delta\frac{\min_{l\in Y[n]}q_2^l[n]}{u_2^{q_1^l[n]}}\right) \right]$
		\State return $\frac{1}{R}\left( \alpha r^1+\beta r^2\right)r^3$
	\end{algorithmic}
\end{algorithm}
\begin{algorithm}[!htbp]
	\caption{ Resource allocation algorithm} % \cite{Strohmer2006}}
	\label{alg:main} 
	\begin{algorithmic}[1]
		\State Initialize main neural network weights $\bfw^0$ and target neural network weights 
		$\bfw^-=\bfw^0$.
		\State Initialize state $\bfs_0 \in \clS$ and random action $a_0 \in \clA$
		\For{$e=1,2,\dots,E $}\, $E$: number of episodes
			\For{$i=1,2,\dots, I$}\, $I$: number of steps per episode
				\State {sample a point $p$ from $\text{Uniform}\left[ 0,1\right]$}
				\If {$p < \epsilon$}
					\State choose a random action $a_{i}$ uniformly from $\clA$
					\Else
					\State {choose the action as $a_{i}=\argmax_a Q(\bfs_i,a,\bfw)$, where $Q(\bfs_i,a,\bfw)$ is the output of the neural network}
				\EndIf
				\State {take the action $a_i$, receive the reward $r_{i}$ according to Alg.\,\ref{alg:rewards}, and observe the next state $\bfs_{i+1}$ from the environment}
				\State{push the tuple $\left\lbrace \bfs_i, a_i, r_i, \bfs_{i+1}, \nu_i\right\rbrace $ into the replay memory}
				\If {replay memory has enough elements}
					\State{
					sample a minibatch of size $M$ as $\left\lbrace \bfs_j, a_j, r_j, \bfs_{j+1}, \nu_j\right\rbrace_{j=1}^M $  from the replay memory}
					\State{
					set the target for every sample of the minibatch according to \eqref{eq:target} as follows   
					\If {$\nu_j == \text{False}$}
					\State{$y_j = r_j+\gamma\max_{a}\hat{Q}(\bfs_{j+1},a,\bfw^-)
					$}
					\Else
						\State{$y_j = r_j
						$}
				
					\EndIf	}
				\State{update the \textit{main} network weights $\bfw$ by performing a gradient descent on  $\sum_j\left\|y_j-Q(s_j,a_j,\bfw)\right\| ^2$}
				\State{update the \textit{target} network weights $\bf\hat{w}$ once every $N_t$ iterations as $\bf\hat{w}\leftarrow\bfw$}
				
				\EndIf
			\EndFor
		\EndFor
	\end{algorithmic}
\end{algorithm}
\section{Simulation Results}
\label{sec:results}
In this section, the details of the parameters used in the simulations are given. Then, the results are presented and discussed.
\subsection{Parameters of the Channel Model} 
Based on the discussion given in Section \ref{sec:model}, the channel vector from the BS to the user corresponding to the $j$th request in the requests' buffer is denoted by $\bfh^j[n]$. This is modeled as
\begin{equation}
\bfh^j[n]=[h^j_1[n], h^j_2[n], \dots, h^j_R[n]]^\transp = \sqrt{l^j}[\zeta^j_1[n], \zeta^j_2[n], \dots, \zeta^j_R[n]]^\transp=\sqrt{l^j}\Phim^{1/2}\bfz^j[n],
\end{equation}
where $l^j$ denotes the large scale fading coefficient, $\Phim$ is the covariance matrix of $\boldsymbol\zeta^j[n] = [\zeta^j_1[n], \zeta^j_2[n], \dots, \zeta^j_R[n]]^\transp$ with the entries given as  $\left[ \bf \Phi\right] _{m,l}=\omega^{\left| m-l\right| }$ for some correlation parameter $\omega \in [0,1]$ and $\bfz^j[n] \sim \cg\left( 0,\bfI_R\right)$. Large scale fading coefficient $l^j$ is assumed to be constant as long as the $j$th request is in the requests' buffer and is also equal for all different RBs. $\bfz^j[n]$ is re-sampled once every coherence period to emphasize that the small scale fading coefficient vector $\boldsymbol\zeta^j[n]$ remains unchanged throughout each coherence period and changes independently from one coherence period to another.
 Large scale fading coefficient and correlation parameter $\omega$ are chosen based on Table \ref{tab:large_scale} \cite{Goldmsith_wls}. As can be seen from Table \ref{tab:large_scale}, users are assumed to be uniformly distributed at a distance between 10 and 100 meters from the BS, which operates with a transmit power of $100\, \text{mW}$ \cite{small_cell14}. Coherence time for small scale fading coefficients is taken to be $12\, \text{ms}$\footnote{Corresponding to a user with a velocity of $10\, \text{m/s}$ with the approximate formula $\frac{0.4}{v/c\times f_c}$ for the coherence time \cite{rappaport}.}.
White noise is considered on each RB with a bandwidth equal to that of an RB ($W$),  at an ambient temperature of $300\, \text{K}$. Each UE's receiver is assumed to have a noise figure of $9$ dB.\\
As discussed in Section \ref{sec:model}, each UE returns a function of its estimated SINR, called the CQI, to the BS. In LTE, CQI is a 4-bit integer. As the relation between CQI and SINR is vendor-specific, we use a simple lookup table to map the estimated SINR to CQI and spectral efficiency. The mapping is given in \cite{LTE_MATLAB13}.\\
% \footnote{Note that the relation between CQI and \SINR\ is not the focus of this paper. It is used just to evaluate the performance of the proposed method for resource allocation.}.\\
As explained in Section \ref{sec:model}, resource allocation is performed on an RB basis. Each RB is assumed to have a time-width of  $T=1\,\text{ms}$ and a frequency-width of $W=180\,\text{KHz}$\footnote{These values are set to be the same as those of the LTE standard, which are 180 KHz for bandwidth and 1 ms for time-width of a single resource block \cite{LTE_resource14, pkt_schedule_LTE13}.}. The number of available RBs at the BS is assumed to be $R=6$.
\begin{table}
	\caption{Channel model parameters}
	\label{tab:large_scale}       % Give a unique label
	\begin{tabularx}{1\columnwidth}{lll}
		\hline\noalign{\smallskip}
		Parameter & Value & description\\
		\noalign{\smallskip}\hline\noalign{\smallskip}	
		$l^j [dB]$
		& \( K - 10\eta{\log _{10}}({\frac{d^j}{d_0}}) + {X^j}\)& large scale fading coefficient between the $j$the user and the BS  \\
		\(d^j [m]\) & $\text{Uniform} \left( 10, 100\right) $ & distance between \textit{i}th user and the BS \\
		\({X^j} \) &  \({\mathcal{N}}\left( {0,\sigma _{sh}^2} \right)\) &shadowing effect\\
		\({\sigma _{sh}}\) & 5.2 &shadowing standard deviation\\
		$K$ &  $20\log_{10}\frac{c}{4\pi d_0f_c}$ &free space path loss\\
		$d_0 [m]$ &  $10$  &reference distance\\
		$f_c$ [GHz]& $1$ &carrier frequency\\
		$\eta$ & $3.5$ &path loss exponent\\
		$\omega$ & $0.001$ & small scale fading correlation parameter\\
		\noalign{\smallskip}\hline
	\end{tabularx}
\end{table}
%\begin{table}
%	% table caption is above the table
%	\caption{Lookup table for mapping the estimated SINR to MCS and SE}
%	\label{tab:MCS_map}       % Give a unique label
%	% For LaTeX tables use
%	\begin{tabularx}{1\columnwidth}{lllll}
%		\hline\noalign{\smallskip}
%		CQI & Modulation& Coding rate& Spectral efficiency [b/s/Hz]& SINR estimate [dB] \\
%		\noalign{\smallskip}\hline\noalign{\smallskip}
%		0&'out of range'&--&--&$<-6.7$\\	
%		1&QPSK&0.0762&0.1523&-6.7\\
%		2&QPSK&0.1172&0.2344&-4.7\\
%		3&QPSK&0.1885&0.3770&-2.3\\
%		4&QPSK&0.3008&0.6016&0.2\\
%		5&QPSK&0.4385&0.8770&2.4\\
%		6&QPSK&0.5879&1.1758&4.3\\
%		7&16QAM&0.3691&1.4766&5.9\\
%		8&16QAM&0.4785&1.9141&8.1\\
%		9&16QAM&0.6016&2.4063&10.3\\
%		10&64QAM&0.4551&2.7305&11.7\\
%		11&64QAM&0.5537&3.3223&14.1\\
%		12&64QAM&0.6504&3.9023&16.3\\
%		13&64QAM&0.7539&4.5234&18.7\\
%		14&64QAM&0.8525&5.1152&21.0\\
%		15&64QAM&0.9258&5.5547&22.7\\
%		
%		\noalign{\smallskip}\hline
%	\end{tabularx}
%\end{table}
\subsection{Request Generation}
\label{sub:req_gen}
For the purpose of simulation, we assume that the arrival process of the requests is Poisson and hence, the inter-arrival time of every two consecutive requests is exponential \cite{papoulis_stochastic}. Three different service types are considered according to Table \ref{tab:rq_dist} \footnote{Parameters for type 1 and type 2 services are chosen as to correspond to typical real-time audio and video applications, respectively. Type 3 corresponds to more delay-tolerant applications with larger PDUs such as video streaming.}. Each service type has a specific Protocol Data Unit (PDU) size, a maximum tolerable latency and also a specific frequency of generation. The last two columns specify the mean of the inter-arrival times for the requests of each service type.
\begin{table}
	\caption{Requests' distributions parameters}
	\label{tab:rq_dist}       % Give a unique label
	\begin{tabularx}{1\columnwidth}{XXXXX}
		\hline\noalign{\smallskip}
		service type & PDU size [Kbits] & maximum tolerable latency [ms] &\multicolumn{2}{c}{mean of the inter-arrival time [ms]}\\
		\cline{4-5}
		&&&low rate&high rate\\
		\noalign{\smallskip}\hline\noalign{\smallskip}	
		type 1 & 3.2 & 150 & 10&5 \\
		type 2 & 64 & 200 & 50&25\\
		type 3 & 200 & 300 & 100&50\\
		\noalign{\smallskip}\hline
	\end{tabularx}
\end{table}
The higher the mean of a request type, the less that type is received by the BS. So, type\,1 has the highest frequency of generation and type\,3 has the lowest. The length of the BS's requests' buffer is set to $L=10$, unless otherwise stated.
%{\small
%	\bibliographystyle{IEEEtran}
%	\bibliography{references}
%	%\bibitem{Bjornson}
%}
\subsection{Parameters of the Neural Network }
Fully connected layers were used to build the neural network used in the simulations. The detailed parameters are given in Table \ref{tab:neural_net}.
\subsection{RL Parameters}
In order to train the neural network, several episodes are used where in each, a fixed number of RL steps are considered. For each episode, a set of requests are  sampled according to the distribution explained in Section \ref{sub:req_gen}. The number of episodes and the number of RL steps in each episode are given in Table \ref{tab:neural_net}. The value of $\epsilon[i]$ for each RL step $i$, used in the $\epsilon$-greedy strategy explained in Section \ref{sub:epsilon_greedy}, is determined based on the following expression \cite{epsilon_greedy11}
\begin{equation}
\epsilon[i]=\epsilon[i-1]-\frac{\epsilon_0-\epsilon_\infty}{\rho},\ \forall i\geq0
\end{equation}
where the parameters $\epsilon_0$, $\epsilon_\infty$, and $\rho$ are given in Table \ref{tab:neural_net}.\\
In order to fairly evaluate the proposed scheme against other methods, we consider two sets of iterations of the same size. The agent is mostly trained during the first set, as $\epsilon[i]$ approaches $\epsilon_\infty$ by the end of the first set. During the second set, the evaluation parameters are gathered while $\epsilon[i]$ remains equal to $\epsilon_\infty$. As $\epsilon_\infty\neq0$, the agent is still able to learn during the second set, but the exploration is marginal. Unless otherwise stated, for each set, the number of episodes and the number of RL steps per episode are chosen based on Table \ref{tab:neural_net}.
%Reward parameters that have been defined earlier in Section\Red{Section??} \ref{sub:rewards} are also given in Table \ref{tab:reward_params}
%Common techniques used in neural nets such as \textit{dropout} and \textit{$l_2$  weight normalization} were used. The former is simply removing a fraction of each layers' weights during training to prevent overfitness 
\subsection{Evaluation Metrics}
In order to measure the capability of the proposed scheme in serving both the licensed and the unlicensed users simultaneously, the following scenario is considered. For the licensed users, the BS receives the requests through a process that was detailed in Section \ref{sec:model}. At the same time, unlicensed users try to use the spectrum vacancies to communicate. This has the benefit of increasing the overall spectral efficiency of the system, provided that the licensed users' performance is not degraded. As sensing the channel to find vacancies has some overhead, such as the mechanisms used in channel access like CSMA/CA \cite{computer_net}, unlicensed users are only able to effectively use the vacancies if the continuity of the vacancy at the same RB exceeds some value (refer to Definition \ref{def:cont_vec}).\\
 For unlicensed users, unlike the licensed ones, we do not consider a QoS-aware scheme, as detailed in previous sections. We only consider a single link consisting of a single transmitter and a single receiver. The link channel parameters are exactly the same as those of any of the links between the BS and any licensed user. In summary, for each coherence period, the receiver is put at a random distance to the transmitter with the channel parameters given in Table \ref{tab:large_scale} and Section \ref{subsec:ch_model}. The spectral efficiency for the unlicensed users becomes nonzero as soon as the continuity function, Definition \ref{def:cont_func}, is equal to $1$. This can be calculated as 
\begin{equation}
\text{SE}_{\text{unlicensed}} = \frac{b_T}{WTN_{\text{unlicensed}}},
\end{equation}
where $b_T$ denotes the total number of bits delivered on the unlicensed link and $N_{\text{unlicensed}}$ denotes the total number of unallocated RBs that satisfy the continuity function's constraint, i.e. produce $g^k_C[n]=1$.\\
Inspired by \eqref{eq:objective}, we use two different notations to evaluate spectral efficiency for the licensed users, namely $\text{SE}_{\text{licensed}}$ and $\widetilde{\text{SE}}_{\text{licensed}}$, where the former refers to the average spectral efficiency including the allocated bits for the missed requests while in the latter those bits are removed.\\
The performance of the proposed scheme is measured against that of the two scheduling methods, namely Maximum Throughput (MT) and minimum Latency (mL). The former allocates each RB to the request whose user has the highest spectral efficiency, i.e. $\hat{a_i}=\argmax_{a_i}\text{SE}(\bfs_i,a_i)$, and the latter chooses the request with the least normalized TTL, i.e. $\hat{a_i}=\argmin_{l\in Y[i]}\left( \frac{q_2^l[i]}{u_2^{q_1^l[i]}}\right) $. As these two methods have no control over the continuity of unallocated resources, we give a fraction of the whole bandwidth, in terms of the number of total RBs, to the licensed users and the rest to unlicensed users to be able to compare them with the proposed scheme. We call these two schemes \lq MT+F\rq and \lq mL+F\rq, respectively.\\
Finally, we define the \textit{acceptance ratio} as the ratio of the number of accepted requests, not dropped, to the total number of arrived requests and \textit{missed ratio} as the ratio of the number of missed requests to the number of accepted requests.
\begin{table}
	\caption{Neural network and RL parameters}
	\label{tab:neural_net}       % Give a unique label
	\begin{tabularx}{1\columnwidth}{llll}
		\hline\noalign{\smallskip}
		parameter & value & parameter & value \\
		\noalign{\smallskip}\hline\noalign{\smallskip}	
		layers' type & Fully Connected&time steps per episode $(I)$& 500 \\
		number of input nodes & $\left( R\!+\!3\right)L\!+\!R\!+\!1\!=\!97$&replay memory size & 100000\\
		number of output nodes & L+1=11&minimum number of observations before training & 1000\\
		number of hidden layers & 3& minibatch size $(M)$ & 32\\
		number of nodes for hidden layers & $[512, 512, 512]$&target network update frequency (in RL steps) $(N_t)$ & 100\\
		weight initialization & $\mathcal{N}\left( 0, 0.05\right)$&$\epsilon_0$& 1 \\
		learning rate $(\kappa)$ & 0.0001&$\epsilon_\infty$&0.01\\
		number of episodes $(E)$& 133&$\rho$&80000\\
		\noalign{\smallskip}\hline
	\end{tabularx}
\end{table}

%\begin{table}
%	% table caption is above the table
%	\caption{RL parameters}
%	\label{tab:RL_params}       % Give a unique label
%	% For LaTeX tables use
%	\begin{tabularx}{1\columnwidth}{ll}
%		\hline\noalign{\smallskip}
%		parameter & value \\
%		\noalign{\smallskip}\hline\noalign{\smallskip}	
%		
%		
%		\\
%		\\
%		\\ 
%		\\
%		\\
%		\\
%		\\
%		\noalign{\smallskip}\hline
%	\end{tabularx}
%\end{table}

%\begin{table}
%	% table caption is above the table
%	\caption{reward parameters}
%	\label{tab:reward_params}       % Give a unique label
%	% For LaTeX tables use
%	\begin{tabularx}{1\columnwidth}{lll}
%		\hline\noalign{\smallskip}
%		parameter & continuity constraint in \eqref{eq:max_objective_simple} in place&continuity constraint in \eqref{eq:max_objective_simple} dropped \\
%		\noalign{\smallskip}\hline\noalign{\smallskip}	
%		$r_e$&0&0\\
%		$r_i$&0&-1\\
%		$r_z$&$-1/2$&-1\\
%		$r_c$&\Red{number of successive unallocted RBs}&-1\\
%		$r_{c_{max}}$&20&--\\
%		$r_T$&0&-1\\
%		$\beta$&0.35&1\\
%		\noalign{\smallskip}\hline
%	\end{tabularx}
%\end{table}
\subsection{Numerical Results}
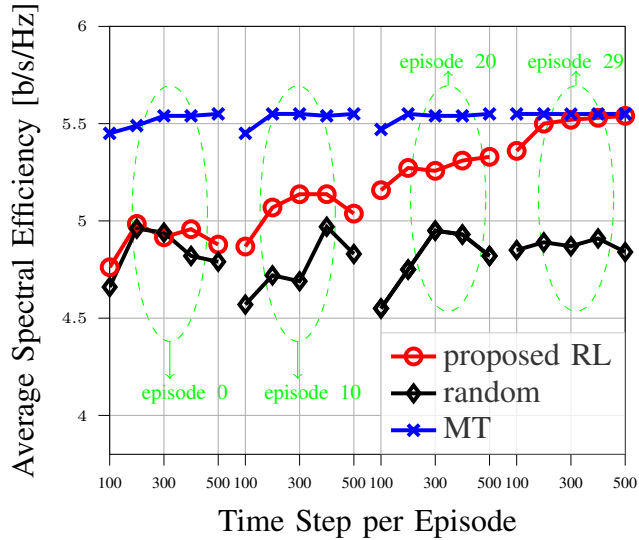
\begin{figure*}
 	\centering
 	% This file was created by tikzplotlib v0.9.1.
\begin{tikzpicture}
%
%\definecolor{color0}{rgb}{0.12156862745098,0.466666666666667,0.705882352941177}
%\definecolor{color1}{rgb}{1,0.498039215686275,0.0549019607843137}
%\definecolor{color2}{rgb}{0.172549019607843,0.627450980392157,0.172549019607843}
%\definecolor{color3}{rgb}{0.83921568627451,0.152941176470588,0.156862745098039}
%\definecolor{color4}{rgb}{0.580392156862745,0.403921568627451,0.741176470588235}
%\definecolor{color5}{rgb}{0.549019607843137,0.337254901960784,0.294117647058824}
%\definecolor{color6}{rgb}{0.890196078431372,0.466666666666667,0.76078431372549}
%\definecolor{color7}{rgb}{0.737254901960784,0.741176470588235,0.133333333333333}
%\definecolor{color8}{rgb}{0.0901960784313725,0.745098039215686,0.811764705882353}
%\draw[red,ultra thick,rounded corners] (7.5,5.3) rectangle (9.4,6.2);
%%%%%%%%%%%%%%%%% ellipses
\draw [green, dashed](0.8,3.2) ellipse (0.5cm and 1.7cm);
\draw [->, green] (0.8, 1.5)-- (0.8, 1);
\node[font=\scriptsize, green] at  (1, 0.8){episode\, 0};

\draw [green, dashed](2.5,3.2) ellipse (0.5cm and 1.7cm);
\draw [->, green] (2.5, 1.5)-- (2.5, 1);
\node[font=\scriptsize, green] at  (2.7, 0.8){episode\, 10};

\draw [green, dashed](4.5,3.4) ellipse (0.5cm and 1.5cm);
\draw [->, green] (4.5, 4.9)-- (4.5, 5.1);
\node[font=\scriptsize, green] at  (4.5, 5.2){episode\, 20};

\draw [green, dashed](6.2,3.4) ellipse (0.5cm and 1.5cm);
\draw [->, green] (6.2, 4.9)-- (6.2, 5.1);
\node[font=\scriptsize, green] at  (6.2, 5.2){episode\, 29};

\begin{axis}[
legend cell align={left},
legend style={fill opacity=0.8, draw opacity=1, text opacity=1, at={(1, 0)}, anchor=south east, draw=white!80!black},
tick align=outside,
tick pos=left,
x grid style={white!69.0196078431373!black},
xlabel={Time Step per Episode},
xmajorgrids,
xmin=100, xmax=2000,
xtick style={color=black},
ticklabel style={font=\tiny},
y grid style={white!69.0196078431373!black},
ylabel={Average Spectral Efficiency [b/s/Hz]},
ymajorgrids,
ymin=3.8, ymax=6,
ytick style={color=black},
xtick={100, 300, 500, 600, 800, 1000, 1100, 1300, 1500, 1600, 1800, 2000}, xticklabels={100, 300, 500, 100, 300, 500, 100, 300, 500, 100, 300, 500}
]
\addplot [line width=1.5pt, mark=o, mark options={solid}, color=red , mark size=3pt, forget plot]
table {%
100 4.7617
200 4.9846
300 4.9158
400 4.9576
500 4.8783
};
%\addlegendentry{episode\,0}

\addplot [line width=1.5pt, mark=o, mark options={solid}, color=red , mark size=3pt, forget plot]
table {%
600 4.8683
700 5.0689
800 5.1370
900 5.1374
1000 5.0375
};
%\addlegendentry{episode\,10}

\addplot [line width=1.5pt, mark=o, mark options={solid}, color=red , mark size=3pt, forget plot]
table {%
	1100 5.1579
	1200 5.2723
	1300 5.2574
	1400 5.3101
	1500 5.3294
};
%\addlegendentry{episode\,20}

\addplot [line width=1.5pt, mark=o, mark options={solid}, color=red , mark size=3pt]
table {%
	1600 5.36
	1700 5.50
	1800 5.52
	1900 5.53
	2000 5.54
	
};
%\addlegendentry{episode\,26}
\addlegendentry{proposed RL}
%%%%%%%%%%%%%%%%%%%% random
\addplot [line width=1.5pt, mark={diamond}, mark options={solid}, color=black , mark size=3pt, forget plot]
table {%
	100 4.66
	200 4.96
	300 4.94
	400 4.82
	500 4.79
};

\addplot [line width=1.5pt, mark={diamond}, mark options={solid}, color=black , mark size=3pt, forget plot]
table {%
	600 4.57
	700 4.72
	800 4.69
	900 4.97
	1000 4.83
};

\addplot [line width=1.5pt, mark={diamond}, mark options={solid}, color=black , mark size=3pt, forget plot]
table {%
	1100 4.55
	1200 4.75
	1300 4.95
	1400 4.93
	1500 4.82
};
\addplot [line width=1.5pt, mark={diamond}, mark options={solid}, color=black , mark size=3pt]
table {%
	1100 4.55
	1200 4.75
	1300 4.95
	1400 4.93
	1500 4.82
};

\addplot [line width=1.5pt, mark={diamond}, mark options={solid}, color=black , mark size=3pt, forget plot]
table {%
	1600 4.85
	1700 4.89
	1800 4.87
	1900 4.91
	2000 4.84
};
\addlegendentry{random}

%%%%%%%%%%%%%%%%%%%% exhaustive
\addplot [line width=1.5pt, mark=x, mark options={solid}, color=blue , mark size=3pt, forget plot]
table {%
	100 5.45
	200 5.49
	300 5.54
	400 5.54
	500 5.55
};

\addplot [line width=1.5pt, mark=x, mark options={solid}, color=blue , mark size=3pt, forget plot]
table {%
	600 5.45
	700 5.55
	800 5.55
	900 5.54
	1000 5.55
};

\addplot [line width=1.5pt, mark=x, mark options={solid}, color=blue , mark size=3pt, forget plot]
table {%
	1100 5.47
	1200 5.55
	1300 5.54
	1400 5.54
	1500 5.55
};

\addplot [line width=1.5pt, mark=x, mark options={solid}, color=blue , mark size=3pt]
table {%
	1600 5.5499
	1700 5.5499
	1800 5.5489
	1900 5.5499
	2000 5.5499
};
\addlegendentry{MT}

\end{axis}

\end{tikzpicture}
 	% 	\centering
 	\caption{performance of three different algorithms, namely the \lq proposed deep RL algorithm\rq, \lq random resource assignment\rq\ and \lq MT\rq\ in terms of average spectral efficiency (over a window of size 1000 RL steps) vs. time steps per episode $\left\lbrace 100, 200, 300, 400, 500\right\rbrace $ for episodes $\left\lbrace 0, 10, 20, 29\right\rbrace $. For the proposed scheme the parameters are chosen as $\alpha=1$, $\beta=0$, and $\delta=\infty$.}
 	\label{fig:SE_high}
 \end{figure*}

\begin{figure}
	\centering
	\input{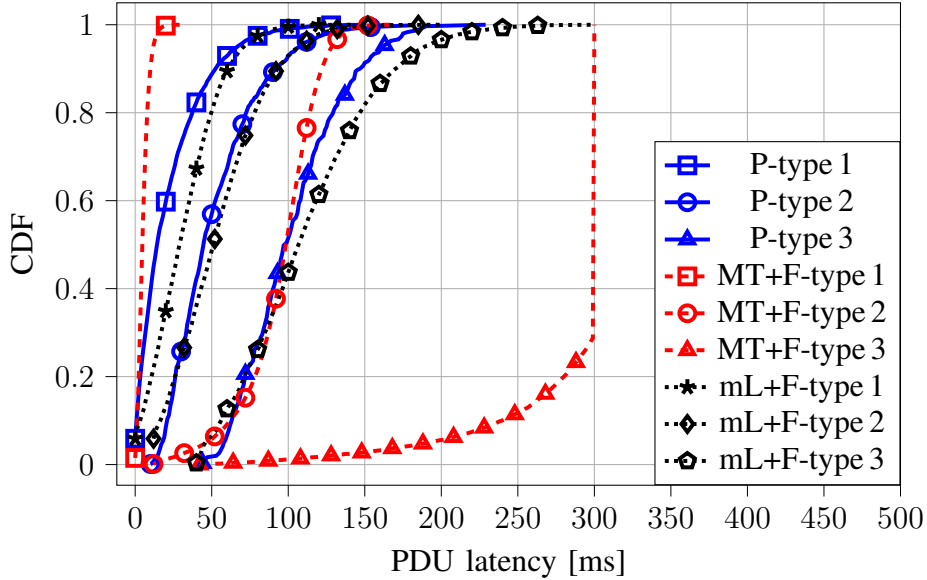}
	\caption{CDF of the aggregated latency of the delivered and missed PDUs for the proposed method (\lq P\rq) (with $\alpha=\beta=2, \delta=1$, $C=2$) against that of the \lq MT+F\rq and \lq mL+F\rq\ methods for different service types in high arrival rate (for all cases $L=10$). The missed ratio for  \lq P\rq, \lq MT+F\rq and \lq ml+F\rq\ methods are $\{2\mathrm{e}{-4}, 5\mathrm{e}{-2}, 2.5\mathrm{e}{-5}\}$, respectively.}
	\label{fig:latency_CDF}
\end{figure}

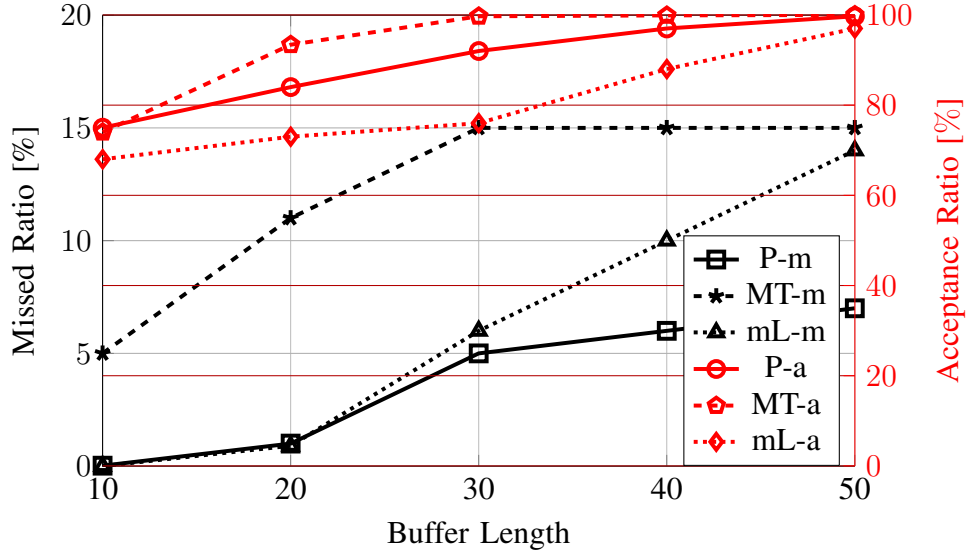
\begin{figure}
	\centering
	%\usepackage{pgfplots}
%\begin{tikzpicture}
%\pgfplotsset{
%	scale only axis,
%	scaled x ticks=base 10:3,
%	xmin=0, xmax5=0
%}
%\begin{axis}[
%axis y line*=right,
%legend style={at={(1, 0)}, anchor=south east},
%tick align=outside,
%tick pos=left,
%x grid style={white!69.0196078431373!black},
%xmajorgrids,
%xmin=10, xmax=50,
%xtick style={color=black},
%y grid style={white!69.0196078431373!black},
%ymajorgrids,
%ymin=-0.0469653179190751, ymax=0.2,
%ylabel={Performance},
%xlabel={Buffer Length},
%ytick style={color=black}
%]
%\addplot [line width=2pt ,mark=square, mark options={solid}, color=green , mark size=3pt ]
%table {%
%	10 0
%	20 0.01
%	30 0.05
%	40 0.06
%	50 0.07	
%};
%\end{axis}
%
%\begin{axis}[
%axis y line*=right,
%ymin=0, ymax=80,
%%xlabel=x-axis,
%ylabel=y-axis 1,
%]
%\addplot[smooth,mark=x,red]
%coordinates{
%	(10,68.6)
%	(20,72)
%	(30,68.6)
%	(40,53.4)
%	(50,22.8)
%}; \label{plot_one}
%\end{axis}
%\end{tikzpicture}

\begin{tikzpicture}
\pgfplotsset{
	scale only axis,
	xmin=10, xmax=50,
	y axis style/.style={
		yticklabel style=#1,
		ylabel style=#1,
		y axis line style=#1,
		ytick style=#1
	}
}
%\pgfplotsset{
%	scale only axis,
%	xmin=10, xmax=50
%}

\begin{axis}[
axis y line*=left,
ymin=0, ymax=20,
xlabel=Buffer Length,
ylabel={Missed Ratio [\%]},
xmajorgrids,
y grid style={white!69.0196078431373!black},
ymajorgrids,
legend style={at={(0.8, 0)}, anchor=south east},
width=10cm,
height=6cm,
xtick={10, 20, 30, 40, 50}, xticklabels={10, 20, 30, 40, 50}
]
%%%%%%%%%%%%%%%% missed ratio
%%%%%%%%%% proposed
\addplot[line width=1.5pt, mark=square,color=black, mark size=3pt]
table {%
		10 0.02
		20 1
		30 5
		40 6
		50 7	
	}; \label{missed-P}
%%%%%%%%%%% MT
\addplot[dashed, line width=1.5pt, mark=star, color=black, mark options=solid,mark size=3pt]
table{
10 5	
20 11
30 15
40 15	
50 15
};\label{missed-MT}
%%%%%%%%%% min delay
\addplot[dotted, line width=1.5pt, mark=triangle, color=black, mark options=solid,mark size=3pt]
table{
10 0.0025
20 0.9
30 6
40 10
50 14

};\label{missed-minD}

\end{axis}

%%%%%%%%%%%%%%%%% acceptance rate
\begin{axis}[
axis y line*=right,
axis x line=none,
ymin=0, ymax=100,
ylabel={Acceptance Ratio [$\%$]},
y axis style=red!99!black,
xmajorgrids,
y grid style={red!70!black},
ymajorgrids,
legend style={at={(0.98, 0)}, anchor=south east},
width=10cm,
height=6cm,
]
\addlegendimage{/pgfplots/refstyle=missed-P}\addlegendentry{P-m}
\addlegendimage{/pgfplots/refstyle=missed-MT}\addlegendentry{MT-m}
\addlegendimage{/pgfplots/refstyle=missed-minD}\addlegendentry{mL-m}
%%%%%%%%% proposed
\addplot[line width=1.5pt,mark=o,red,mark size=3pt]
table {%
	10 75
	20 84
	30 92
	40 97
	50 99.7	
};\label{acc-P}
%\addlegendentry{P-acceptance}
%%%%%%%% MT
\addplot[dashed,line width=1.5pt, mark=pentagon,red,mark size=3pt, mark options=solid]
table{
10 74
20 93.4
30 99.7
40 99.9
50 100
};\label{acc-MT}
%%%%%% min delay
\addplot[dotted,line width=1.5pt, mark=diamond, color=red, mark options=solid,mark size=3pt]
table{
	10 68
	20 73
	30 76
	40 88
	50 97
};\label{acc-minD}
%\addlegendentry{MT-acceptance}

\addlegendimage{/pgfplots/refstyle=acc-P}\addlegendentry{P-a}
\addlegendimage{/pgfplots/refstyle=acc-MT}\addlegendentry{MT-a}
\addlegendimage{/pgfplots/refstyle=acc-minD}\addlegendentry{mL-a}

\end{axis}

\end{tikzpicture}
	\caption{Missed ratio (\lq m\rq) on the left axis and acceptance ratio (\lq a\rq) on the right axis for the proposed method (\lq P\rq) (with $\alpha=\beta=2, \delta=1$, $C=2$) against those of the \lq MT+F\rq and \lq mL+F\rq methods versus buffer length ($L$) in high arrival rate (for all cases $L=10$).}
	\label{fig:mis_vs_buflen}
\end{figure}

The numerical results for the proposed deep RL based algorithm are given in this section. First, we present the learning trend of the proposed scheme in terms of $\text{SE}_{\text{licensed}}$ vs. time steps per episodes against that of a random resource allocation scheme  and also of the MT in Fig. \ref{fig:SE_high}, where we used 30 episodes for the simulations in total.
 For the proposed scheme, the parameters are chosen as $\alpha=1, \beta=0$, and $\delta=\infty$. This removes the second and third terms in \eqref{eq:objective}. In other words, the continuity of the unallocated resources and also the latency constraint are both ignored. This makes MT to have the optimum performance in terms of average spectral efficiency.  It can be seen that the proposed RL algorithm starts with a performance similar to that of the random assignment scheme and gradually improves as time steps pass and finally achieves the performance of MT. For the results in Fig. \ref{fig:SE_high}, 30 episodes are taken and the requests are generated according to the \lq high rate\rq\ column in Table \ref{tab:rq_dist}.\\
A comparison between the performance of the proposed method and that of the \lq MT+F\rq\ and \lq mL+F\rq\ methods are depicted in Fig. \ref{fig:SE_vs_cont} versus the continuity length as in Definition \ref{def:cont_func}. It is seen that the proposed method achieves higher average spectral efficiency (both for the licensed and the unlicensed users) than the other two methods, up to a continuity value of $10$ for low arrival rates. It is also seen that increasing the arrival rate (from low to high) causes the proposed method to focus more on the licensed users and hence, $\text{SE}_{\text{unlicensed}}$ decreases accordingly. This is in spite of the fact that we increased $\beta$ and decreased $\alpha$ to strengthen the continuity term in the rewarding mechanism and in turn to maintain $\text{SE}_{\text{unlicensed}}$ in part. We can increase $\beta$ and decrease $\alpha$ even further, but this can cause further reduction in $\widetilde{\text{SE}}_{\text{licensed}}$ as well. The proposed method is able to support the uninterrupted communication of the unlicensed users as long as the QoS of the licensed users are guaranteed. When the traffic of the requests is higher, i.e. in high arrival rates, there is less available free resources. As a result, $\text{SE}_{\text{unlicensed}}$ drops. The number of RBs given to \lq MT+F\rq\ and \lq mL+F\rq\ are chosen in their favor such that the best sum spectral efficiency ($\widetilde{\text{SE}}_{\text{licensed}} + \text{SE}_{\text{unlicensed}}$) is achieved. It is worthwhile to note that \lq mL+F\rq\ can only achieve an acceptance ratio of $88\,\%$ in high arrival rate with $L=40$. A more fair comparison would be to consider it with $L=50$ against the other two methods with $L=40$. This decreases the sum spectral efficiency of \lq mL+F\rq\ to $4.24\,\text{b/s/Hz}$ which is less than that of the proposed method.\\
\begin{figure}
	\centering
	\input{pics/bar_chart_test4_merged.tex}
	\caption{Average spectral efficiency ($\widetilde{\text{SE}}_{\text{licensed}}$, $\text{SE}_{\text{unlicensed}}$) for the proposed method (\lq P\rq) against that of the \lq MT+F\rq\ and \lq mL+F\rq methods versus the continuity length ($C$) (Definition \ref{def:cont_func}) for two different arrival rates (for all cases $L=40$): Low (\lq l\rq), where $\alpha=2, \beta=2, \delta=1$ for the proposed method. For \lq MT+F\rq\ and \lq mL+F\rq, 4 RBs are given to the licensed and 2 RBs given to the unlicensed users. High (\lq h\rq), where $\alpha=1.5, \beta=2.5, \delta=1$ for the proposed method. For \lq MT+F\rq\ and \lq mL+F\rq, 5 RBs are given to the licensed and 1 RB is given to the unlicensed users. The missed ratio for \lq MT+F\rq and \lq ml+F\rq\ methods are $0.15$ and $0.1$, respectively while at most $0.08$ for the proposed method.
		%			 The gain achieved by the proposed method over the \lq MT+F\rq\ method in terms of the total average spectral efficiency ($\widetilde{\text{SE}_\text{\text{licensed}}}+\text{SE}_{\text{unlicensed}}$)for  $C=\{5, 10, 20\}$ are approximately $\{21\%, 15\%, 1\%\}$, respectively.
	}
	\label{fig:SE_vs_cont}
\end{figure}
In Fig. \ref{fig:latency_CDF}, the CDF of the delivered/missed PDUs of the requests for the proposed method is depicted against that of the \lq MT+F\rq\ and \lq mL+F\rq\ methods. It is seen that the proposed method serves all three types of requests well before their deadlines. In fact it acts, in terms of latency, similar to \lq mL\rq, as it takes into account the maximum tolerable latencies of the requests. \lq MT+F\rq\ method, however, missed many of the requests of type\,3 as it only takes into account the instantaneous spectral efficiency. The missed ratio shows that the proposed method delivers the requests with $2\mathrm{e}{-4}$ missed ratio, while \lq MT+F\rq\ has a missed ratio of $5\mathrm{e}{-2}$.\\
The performance of the three methods vs. the buffer length ($L$) in terms of missed ratio and acceptance ratio are depicted in Fig. \ref{fig:mis_vs_buflen}. It is seen that the proposed method can achieve up to 99.6\,\% acceptance ratio for high arrival rate by increasing buffer length from 10 to 50. It is also seen that the proposed method has the least increase in missed ratio by increasing the buffer length among the three methods and can also accept almost all of the incoming requests with the least missed ratio among all the methods.\\
 $\widetilde{\text{SE}}_{\text{licensed}}$ and $\text{SE}_{\text{unlicensed}}$ are also compared in Fig. \ref{fig:SE_vs_buflen} for low and high arrival rates in terms of buffer length. All the three methods show slight variation, in terms of spectral efficiency, for low arrival rates. This is due to the fact that the buffer with $L=10$ almost suffices to hold all the incoming requests. For high arrival rates, however, spectral efficiencies change with the change of buffer length. As the buffer length increases, more requests are accepted and should be handled in time. The proposed method, as noted earlier, can handle almost all of the requests with the least decrease in sum spectral efficiency. The other two methods fall short in serving all of the incoming requests.
%\begin{figure}
%	\centering
%	\input{pics/SE_vs_cont.tex}
%%	\caption{Average spectral efficiency for the proposed method (\lq P\rq) (\Red{alpha, beta, delta, L}) against that of the \lq MT+F\rq\ method \Red{add mL} (with 4 RBs given to the licensed and 2 RBs given to the unlicensed users) versus the continuity length ($C$) (Definition \ref{def:cont_func}) for low requests' rate (Table \ref{tab:rq_dist}). The gain achieved by the proposed method over the \lq MT+F\rq\ method in terms of the total average spectral efficiency for  $C=\{5, 10, 20\}$ are approximately $\{21\%, 15\%, 1\%\}$, respectively. }
%%	\label{fig:SE_vs_cont}
%\end{figure}
\begin{figure}
	\centering
	\input{pics/SE_vs_buflen2.tex}
	\caption{Average spectral efficiency ($\widetilde{\text{SE}}_{\text{licensed}}$, $\text{SE}_{\text{unlicensed}}$) for the proposed method (\lq P\rq) against that of the \lq MT+F\rq\ and \lq mL+F\rq methods versus buffer length ($L$) with $C=2$ (Definition \ref{def:cont_func}) for two different arrival rates: Low (\lq l\rq), where $\alpha=2, \beta=2, \delta=1$ for the proposed method. For \lq MT+F\rq\ and \lq mL+F\rq, 4 RBs are given to the licensed and 2 RBs given to the unlicensed users. High (\lq h\rq), where $\alpha=1.5, \beta=2.5, \delta=1$ for the proposed method. For \lq MT+F\rq\ and \lq mL+F\rq, 5 RBs are given to the licensed and 1 RB is given to the unlicensed users.}
	\label{fig:SE_vs_buflen}
\end{figure}
\section{Acknowledgment}
The work by  Mahdi Nouri Boroujerdi is supported by Iran National Science Foundation (INSF). The work by Babak Hossein Khalaj, Mohammad Ali Maddah-Ali, Mohammad Akbari and Roghayeh Joda is funded by ICT Research Institute (ITRC). 
\section{Conclusion}
\label{sec:conclusion}
In this paper, we proposed a deep-RL-based resource allocation for a cellular network which can make a balance among spectral efficiency, from the network manager perspective, the quality of service, from the users' perspective, and smoothness of the unallocated resources, which is beneficial to the unlicensed users. Results show that the proposed scheme can efficiently make such balance through the proposed learning mechanism.
{\small
	\bibliographystyle{IEEEtran}
	\bibliography{references}

\begin{thebibliography}{10}
\providecommand{\url}[1]{#1}
\csname url@samestyle\endcsname
\providecommand{\newblock}{\relax}
\providecommand{\bibinfo}[2]{#2}
\providecommand{\BIBentrySTDinterwordspacing}{\spaceskip=0pt\relax}
\providecommand{\BIBentryALTinterwordstretchfactor}{4}
\providecommand{\BIBentryALTinterwordspacing}{\spaceskip=\fontdimen2\font plus
\BIBentryALTinterwordstretchfactor\fontdimen3\font minus
  \fontdimen4\font\relax}
\providecommand{\BIBforeignlanguage}[2]{{%
\expandafter\ifx\csname l@#1\endcsname\relax
\typeout{** WARNING: IEEEtran.bst: No hyphenation pattern has been}%
\typeout{** loaded for the language `#1'. Using the pattern for}%
\typeout{** the default language instead.}%
\else
\language=\csname l@#1\endcsname
\fi
#2}}
\providecommand{\BIBdecl}{\relax}
\BIBdecl

\bibitem{DSS_cognitive_survey}
W.~S. H. M.~W. {Ahmad}, N.~A.~M. {Radzi}, F.~S. {Samidi}, A.~{Ismail},
  F.~{Abdullah}, M.~Z. {Jamaludin}, and M.~N. {Zakaria}, ``{5G} technology:
  Towards dynamic spectrum sharing using cognitive radio networks,'' \emph{IEEE
  Access}, vol.~8, pp. 14\,460--14\,488, 2020.

\bibitem{LTE-UvsLAA_19}
B.~{Bojović}, L.~{Giupponi}, Z.~{Ali}, and M.~{Miozzo}, ``Evaluating
  unlicensed {LTE} technologies: {LAA} vs {LTE-U},'' \emph{IEEE Access},
  vol.~7, pp. 89\,714--89\,751, 2019.

\bibitem{LTE-U_sensing20}
Q.~{Yang}, Y.~{Huang}, Y.~{Yen}, L.~{Chen}, H.~{Chen}, X.~{Hong}, J.~{Shi}, and
  L.~{Wang}, ``Location based joint spectrum sensing and radio resource
  allocation in cognitive radio enabled {LTE-U} systems,'' \emph{IEEE
  Transactions on Vehicular Technology}, vol.~69, no.~3, pp. 2967--2979, 2020.

\bibitem{pkt_schedule_LTE13}
F.~{Capozzi}, G.~{Piro}, L.~A. {Grieco}, G.~{Boggia}, and P.~{Camarda},
  ``Downlink packet scheduling in {LTE} cellular networks: Key design issues
  and a survey,'' \emph{IEEE Communications Surveys Tutorials}, vol.~15, no.~2,
  pp. 678--700, 2013.

\bibitem{QoS-resource19}
J.~{Tan}, S.~{Xiao}, S.~{Han}, Y.~{Liang}, and V.~C.~M. {Leung}, ``Qos-aware
  user association and resource allocation in {LAA-LTE/WiFi} coexistence
  systems,'' \emph{IEEE Transactions on Wireless Communications}, vol.~18,
  no.~4, pp. 2415--2430, April 2019.

\bibitem{resource_LTE_16}
Y.~L. {Lee}, J.~{Loo}, T.~C. {Chuah}, and A.~A. {El-Saleh}, ``Fair resource
  allocation with interference mitigation and resource reuse for {LTE/LTE-A}
  femtocell networks,'' \emph{IEEE Transactions on Vehicular Technology},
  vol.~65, no.~10, pp. 8203--8217, 2016.

\bibitem{ML_Wls17}
C.~{Jiang}, H.~{Zhang}, Y.~{Ren}, Z.~{Han}, K.~{Chen}, and L.~{Hanzo},
  ``Machine learning paradigms for next-generation wireless networks,''
  \emph{IEEE Wireless Communications}, vol.~24, no.~2, pp. 98--105, April 2017.

\bibitem{deep_wls18}
Q.~{Mao}, F.~{Hu}, and Q.~{Hao}, ``Deep learning for intelligent wireless
  networks: A comprehensive survey,'' \emph{IEEE Communications Surveys
  Tutorials}, vol.~20, no.~4, pp. 2595--2621, Fourthquarter 2018.

\bibitem{DRL_survey19}
N.~C. {Luong}, D.~T. {Hoang}, S.~{Gong}, D.~{Niyato}, P.~{Wang}, Y.~{Liang},
  and D.~I. {Kim}, ``Applications of deep reinforcement learning in
  communications and networking: A survey,'' \emph{IEEE Communications Surveys
  Tutorials}, vol.~21, no.~4, pp. 3133--3174, Fourthquarter 2019.

\bibitem{simpleRL_LTE17}
\BIBentryALTinterwordspacing
E.~C. Santos, ``A simple reinforcement learning mechanism for resource
  allocation in {LTE-A} networks with {Markov} decision process and
  {Q}-learning,'' \emph{CoRR}, vol. abs/1709.09312, 2017. [Online]. Available:
  \url{http://arxiv.org/abs/1709.09312}
\BIBentrySTDinterwordspacing

\bibitem{deepmin_DRL15}
\BIBentryALTinterwordspacing
V.~Mnih, K.~Kavukcuoglu, D.~Silver, A.~A. Rusu, J.~Veness, M.~G. Bellemare,
  A.~Graves, M.~Riedmiller, A.~K. Fidjeland, G.~Ostrovski, S.~Petersen,
  C.~Beattie, A.~Sadik, I.~Antonoglou, H.~King, D.~Kumaran, D.~Wierstra,
  S.~Legg, and D.~Hassabis, ``Human-level control through deep reinforcement
  learning,'' \emph{Nature}, vol. 518, no. 7540, pp. 529--533, 2015. [Online].
  Available: \url{https://doi.org/10.1038/nature14236}
\BIBentrySTDinterwordspacing

\bibitem{deepRL_resource18}
R.~{Li}, Z.~{Zhao}, Q.~{Sun}, C.~{I}, C.~{Yang}, X.~{Chen}, M.~{Zhao}, and
  H.~{Zhang}, ``Deep reinforcement learning for resource management in network
  slicing,'' \emph{IEEE Access}, vol.~6, pp. 74\,429--74\,441, 2018.

\bibitem{DRL_hetnet}
N.~{Zhao}, Y.~{Liang}, D.~{Niyato}, Y.~{Pei}, M.~{Wu}, and Y.~{Jiang}, ``Deep
  reinforcement learning for user association and resource allocation in
  heterogeneous cellular networks,'' \emph{IEEE Transactions on Wireless
  Communications}, vol.~18, no.~11, pp. 5141--5152, Nov 2019.

\bibitem{proactive_DRL18}
U.~{Challita}, L.~{Dong}, and W.~{Saad}, ``Proactive resource management for
  {LTE} in unlicensed spectrum: A deep learning perspective,'' \emph{IEEE
  Transactions on Wireless Communications}, vol.~17, no.~7, pp. 4674--4689,
  July 2018.

\bibitem{cognitive_share_aggregate}
W.~{Zhang}, C.~{Wang}, X.~{Ge}, and Y.~{Chen}, ``Enhanced {5G} cognitive radio
  networks based on spectrum sharing and spectrum aggregation,'' \emph{IEEE
  Transactions on Communications}, vol.~66, no.~12, pp. 6304--6316, 2018.

\bibitem{5GNR_Phy}
``5{G}; {NR}; {Physical} layer procedures for data,'' ETSI, Tech. Rep. ETSI TS
  138 214, 2019.

\bibitem{LTE_wiley09}
S.~Sesia, I.~Toufik, and M.~Baker, \emph{LTE, The UMTS Long Term Evolution:
  From Theory to Practice}.\hskip 1em plus 0.5em minus 0.4em\relax Wiley
  Publishing, 2009.

\bibitem{sutton_RL20}
R.~S. Sutton and A.~G. Barto, \emph{Reinforcement Learning: An Introduction},
  2nd~ed.\hskip 1em plus 0.5em minus 0.4em\relax Cambridge, MA, USA: The MIT
  Press, 2020.

\bibitem{szepesvari2010algorithms}
\BIBentryALTinterwordspacing
C.~Szepesv{\'a}ri, \emph{Algorithms for Reinforcement Learning}, ser. Synthesis
  lectures on artificial intelligence and machine learning.\hskip 1em plus
  0.5em minus 0.4em\relax Morgan \& Claypool, 2010. [Online]. Available:
  \url{https://books.google.com/books?id=qwtphfl7U74C}
\BIBentrySTDinterwordspacing

\bibitem{neural_net_universal01}
B.~C. Cs{\'a}ji, ``Approximation with artificial neural networks,'' Master's
  thesis, Faculty of Sciences, Eötvös Loránd University, Hungary, 2001.

\bibitem{neural_net_express17}
\BIBentryALTinterwordspacing
Z.~Lu, H.~Pu, F.~Wang, Z.~Hu, and L.~Wang, ``The expressive power of neural
  networks: A view from the width,'' in \emph{Advances in Neural Information
  Processing Systems 30}, I.~Guyon, U.~V. Luxburg, S.~Bengio, H.~Wallach,
  R.~Fergus, S.~Vishwanathan, and R.~Garnett, Eds.\hskip 1em plus 0.5em minus
  0.4em\relax Curran Associates, Inc., 2017, pp. 6231--6239. [Online].
  Available:
  \url{http://papers.nips.cc/paper/7203-the-expressive-power-of-neural-networks-a-view-from-the-width.pdf}
\BIBentrySTDinterwordspacing

\bibitem{deepmind_RL15}
\BIBentryALTinterwordspacing
V.~Mnih, K.~Kavukcuoglu, D.~Silver, A.~A. Rusu, J.~Veness, M.~G. Bellemare,
  A.~Graves, M.~Riedmiller, A.~K. Fidjeland, G.~Ostrovski, S.~Petersen,
  C.~Beattie, A.~Sadik, I.~Antonoglou, H.~King, D.~Kumaran, D.~Wierstra,
  S.~Legg, and D.~Hassabis, ``Human-level control through deep reinforcement
  learning,'' \emph{Nature}, vol. 518, no. 7540, pp. 529--533, Feb. 2015.
  [Online]. Available: \url{http://dx.doi.org/10.1038/nature14236}
\BIBentrySTDinterwordspacing

\bibitem{Goldmsith_wls}
A.~Goldsmith, \emph{Wireless Communications}, 1st~ed.\hskip 1em plus 0.5em
  minus 0.4em\relax Cambridge University Press, 2005.

\bibitem{small_cell14}
S.~{Chou}, T.~{Chiu}, Y.~{Yu}, and A.~{Pang}, ``Mobile small cell deployment
  for next generation cellular networks,'' in \emph{2014 IEEE Global
  Communications Conference}, 2014, pp. 4852--4857.

\bibitem{rappaport}
T.~S. Rappaport, \emph{Wireless Communications: Principles and Practice},
  2nd~ed.\hskip 1em plus 0.5em minus 0.4em\relax Prentice Hall, 2002.

\bibitem{LTE_MATLAB13}
H.~Zarrinkoub, \emph{Understanding LTE with MATLAB®: From Mathematical
  Modeling to Simulation and Prototyping}.\hskip 1em plus 0.5em minus
  0.4em\relax Wiley Publishing, 01 2013.

\bibitem{LTE_resource14}
Y.~L. {Lee}, T.~C. {Chuah}, J.~{Loo}, and A.~{Vinel}, ``Recent advances in
  radio resource management for heterogeneous {LTE/LTE-A} networks,''
  \emph{IEEE Communications Surveys Tutorials}, vol.~16, no.~4, pp. 2142--2180,
  Fourthquarter 2014.

\bibitem{papoulis_stochastic}
A.~Papoulis and S.~U. Pillai, \emph{Probability, Random Variables, and
  Stochastic Processes}, 4th~ed.\hskip 1em plus 0.5em minus 0.4em\relax Boston:
  McGraw Hill, 2002.

\bibitem{epsilon_greedy11}
M.~Tokic and G.~Palm, ``Value-difference based exploration: Adaptive control
  between epsilon-greedy and softmax,'' in \emph{KI 2011: Advances in
  Artificial Intelligence}, J.~Bach and S.~Edelkamp, Eds.\hskip 1em plus 0.5em
  minus 0.4em\relax Berlin, Heidelberg: Springer Berlin Heidelberg, 2011, pp.
  335--346.

\bibitem{computer_net}
D.~E. Comer, \emph{Computer Networks and Internets}, 6th~ed.\hskip 1em plus
  0.5em minus 0.4em\relax Pearson, 2014.

\end{thebibliography}
	%\bibitem{Bjornson}
}

\end{document}